\documentclass[aps,prb,twocolumn]{revtex4}
\input epsf
\usepackage{psfig}
\usepackage{graphicx}

\begin{document}
\title{Stiff knots 
}
\author{R. Gallotti, O.\ Pierre-Louis}
\affiliation{CNRS/Laboratoire de Spectrom\'etrie Physique,
 Univ. J. Fourier, Grenoble 1,
BP87, F38402 Saint Martin d'H\`eres, France.}
\date{\today}

\begin{abstract}
We report on the geometry and mechanics of knotted 
stiff strings. We discuss both closed and open knots.
Our two main results are:
(i) Their equilibrium energy as well as the 
equilibrium tension for open knots
depend on the type
of knot as the square of the bridge number; 
(ii) Braid localization is found to be
a general feature of stiff strings entanglements,
while angles and knot localization are forbidden.
Moreover, we identify a family of knots
for which the equilibrium shape is a circular braid.
Two other equilibrium shapes are found from Monte Carlo
simulations. These three shapes are confirmed 
by rudimentary experiments.
Our approach is also extended to the problem of the minimization
of the length of a knotted string with a maximum allowed curvature.
\end{abstract}
\pacs{PACS numbers: }
\maketitle

\section{Introduction}

\begin{figure}
\centerline{
\hbox{\psfig{figure=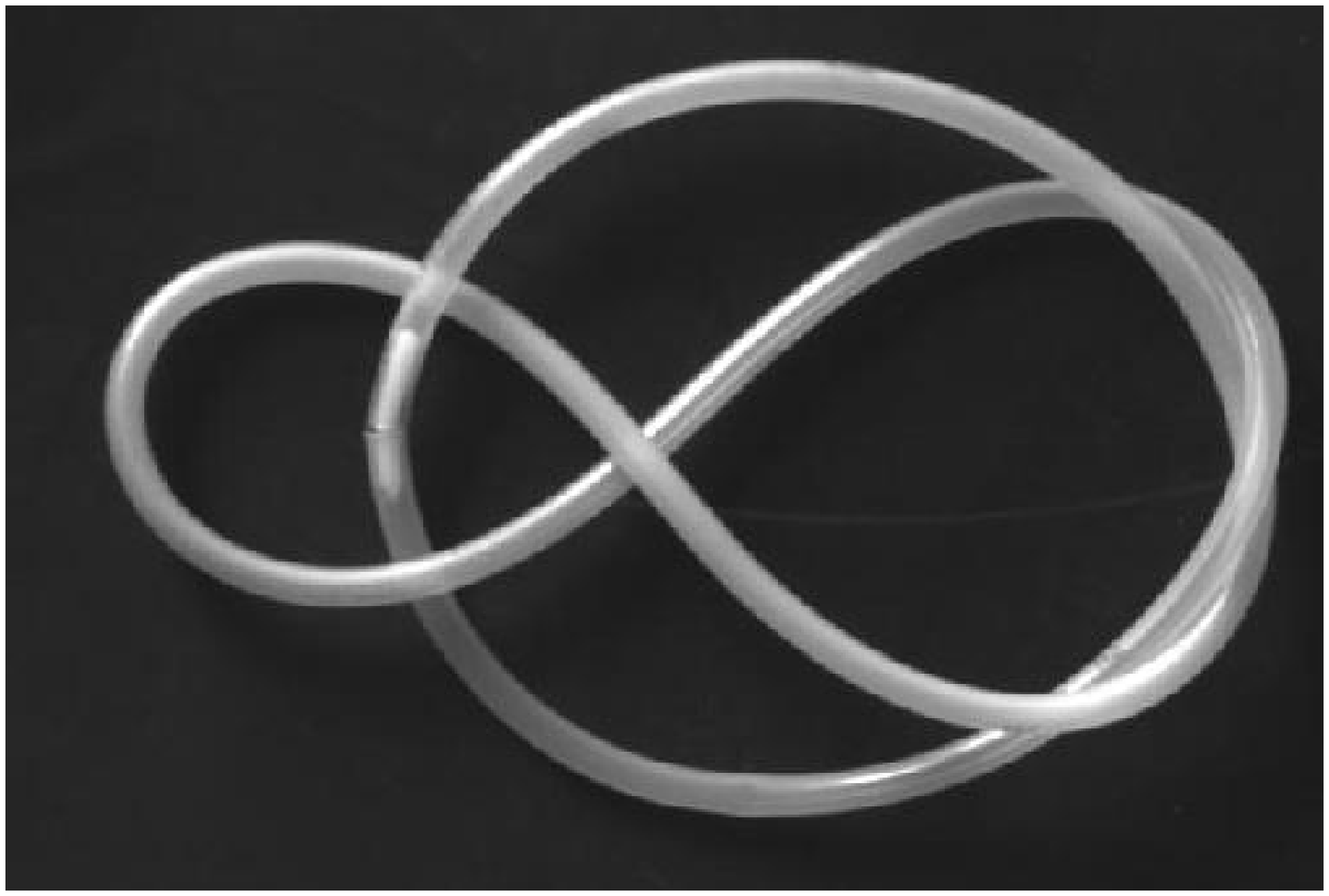,width=5cm,angle=0}}  
}\centerline{
\hbox{\psfig{figure=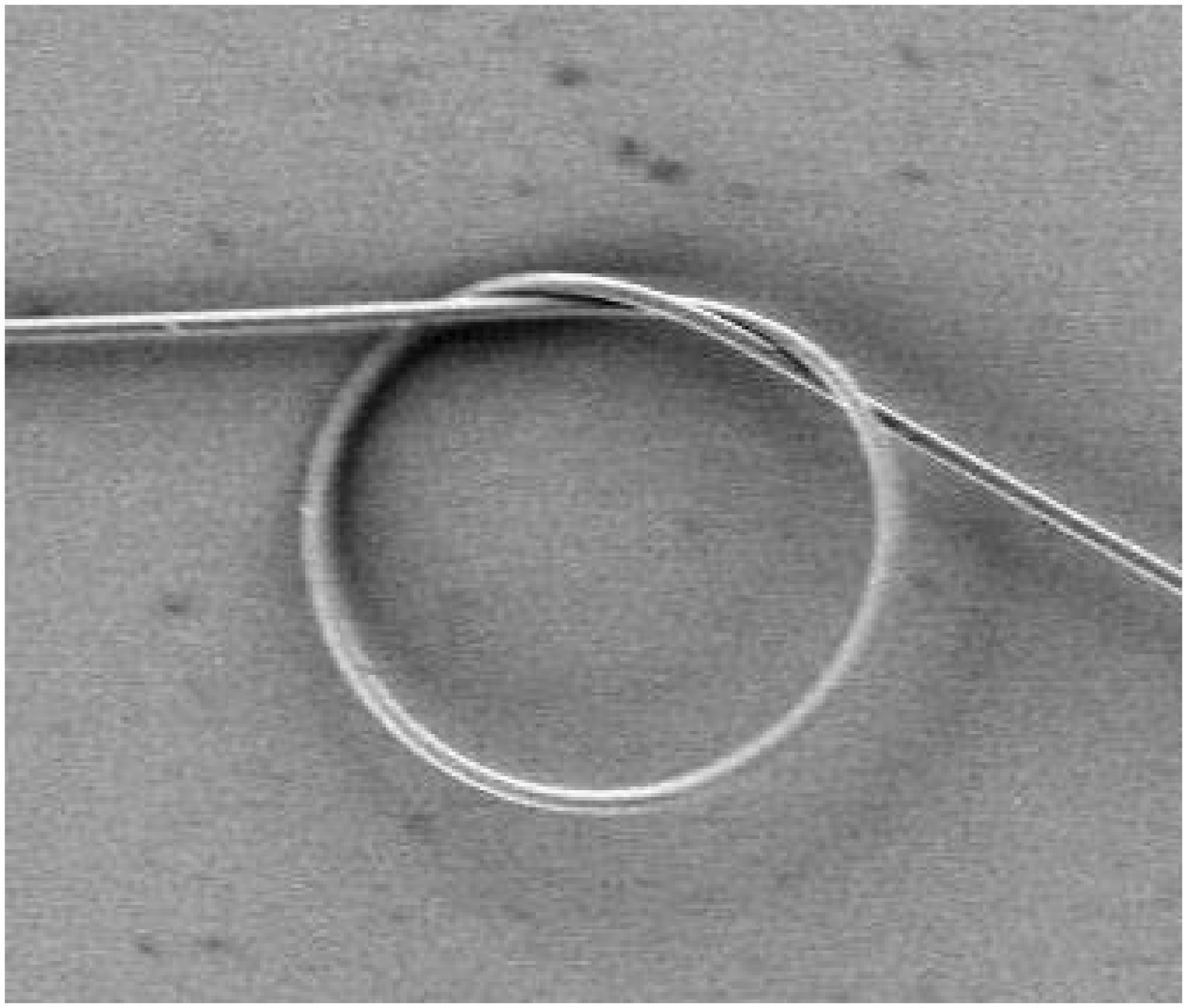,width=5cm,angle=0}}  }
\caption{
Stiff knots. Upper panel: A closed knot,
made with a plastic tube of width
$1$ cm and length $80$ cm
(see section \ref{s:exp} for details). 
Lower panel: Open trefoil knot with a silica nanowire,
from Ref.\cite{Tong2003}. The width of the picture is
20 $\mu$m.
}
\label{fig:phys_knots}
\end{figure}

We report on
the properties of stiff knots, i.e. knotted strings
whose shape is dictated by the bending curvature
energy and contact interactions only. Stiff knots, 
such as loose knots with nylon or metal strings,
are ordinary objects in everyday life (Fig.1).
An upsurge of interest in stiff knots 
recently came  from studies which 
pertain to microscopic objects, such as
those encountered in biology and nano-technologies.
Some examples are:
knots with actin filaments \cite{actin},
nanotubes \cite{Lobovkina2004}, nanotube
fibers \cite{macro}, and silica wires \cite{Tong2003}
(see Fig. \ref{fig:phys_knots}). 
These studies point out the relevance 
of stiff knots for
the experimental determination of the bending rigidity \cite{actin},
for knot induced 
polymer and filament break-up \cite{Saitta1999,actin},
and for  nano-manipulation \cite{Lobovkina2004}.

Moreover, many recent experimental and theoretical
studies were devoted to
knots in polymers, with an emphasis on
knotted DNA \cite{Quake_Vilgis_Katrich}.
They show that
flexible polymers are subject to
knot localisation
leading to the formation of small prime knots.
Knot localization may result from
entropic effects \cite{knot_loc_entro}, or
long range interactions \cite{knot_loc_inter,O'Hara2003}. 
If the localized knots are small enough, their
thermal fluctuations become negligible and
they might be described by the stiff knot regime.

Finally, stiff knots may be considered as elementary
entanglements which capture qualitatively some of the 
features of more complex entanglements. Hence,
 stiff knots may also
provide insights for the curvature 
energy dominated behavior of tightly
entangled semi-flexible polymers such as actin\cite{actin-solutions,
Morse2001}, and other fibrous materials \cite{Rodney2005}.

Here, we aim to establish the basic mechanical
and geometrical properties of stiff knots.
First, the mechanical properties of the knots
are found to exhibit a surprisingly simple 
dependence on  the knot type
via a quantity called the bridge number $n$
(to be defined below). 
We show that the minimum knot energy,
as well as the minimum equilibrium
tension at the open ends of an open knot, 
both increase with $n$ as $n^2$.
Secondly, we identify a striking and general feature 
of the geometry of stiff strings entanglements, 
which we call braid localization.
We analyze the geometries of the simplest knots
which indeed exhibit braid localization.

We will begin with a study of closed knots.
In section II, we define the curvature energy
of a filament. In section III, we discuss the role of interactions
on the equilibrium shape of a filament,
and we place our problem within the historical perspective of the
studies on  the so-called Bernoulli-Euler elastica.
A lower bound for the energy of the equilibrium
configuration is given in section IV.
In section V, we analyze the limit of thin strings 
which leads to braid localization.
In section VI, the results of section V are used 
in order to obtain upper bounds for the equilibrium 
energy. Two results follow from the existence of these
upper bounds: (i) the
global minimum of the energy scales as $n^2$;
(ii) the minimization problem
is solved exactly for a special family of knots. 
The shape and energy of some simple knots
are obtained from Monte Carlo Simulations in section VII.
We find three geometries which 
correspond to the simplest knots. 
In section VIII, we translate the main results of our
analysis to the case of open knots.
In section IX, we discuss several additional points:
(i) the equivalence between the limits of thin strings
and long strings; (ii) the curvature energy of thick knots;
(iii) rudimentary experiments; (iv) the restricted
curvature model, which makes link with the work
of Buck and Rawdon \cite{Buck2004}.
Finally, we conclude in section X.

\section{Model}
\label{s:model}

We shall first
focus on closed knots, i.e. single knotted strings
without free end, as in the upper panel of 
Fig.\ref{fig:phys_knots}. A given configuration of a knot
in 3D space is described by a position vector
${\bf r}(s)$, where $s$ is the arclength.
Since the knot is closed, 
${\bf r}(s)$ is periodic in $s$, and its period is the length
\begin{eqnarray}
{\cal L}=\int ds,
\label{e:length}
\end{eqnarray} 
of the knot. In the following, the absence of bounds
in the integrals indicates integration over the whole knot.
We  define the usual curvature energy of an
inextensible string\cite{Doi,kamien} as:
\begin{eqnarray}
{\cal E}={C \over 2}\int ds \, \kappa^2,
\label{e:E}
\end{eqnarray}
where $\kappa\geq 0$ is the curvature, 
and $C$ is the bending rigidity. Such a modelling  
is valid in the limit of small deformations, defined
by the limit of small curvature \cite{Doi}:
\begin{eqnarray}
w\kappa\ll 1
\label{e:small_deformations}
\end{eqnarray}
where $w$ is the diameter of the section of
the filament.

The question we address is the following:
for a given knot $K$ of length ${\cal L}$,
what are the equilibrium shape and energy?
We call equilibrium energy the global
minimum energy, as opposed to 
that corresponding to possible other local minima,
which will be referred to as metastable states.
The equilibrium energy is denoted as ${\cal E}^*$.
We shall fix ${\cal L}$
by means of a Lagrange multiplier $\mu$.
Mechanical equilibrium is then obtained from
the minimization of:
\begin{eqnarray}
{\cal F}={\cal E}+\mu {\cal L}={ C \over 2}\int ds \, \kappa^2 +\mu \int ds \, .
\label{e:F}
\end{eqnarray}

\section{On the role of interactions}
\label{s:interactions}
Let us first consider this minimization problem
in the absence of interactions between
the different parts of the strings. 
In this case, the curve
can freely cross itself.
We obtain the so-called elastica,
initially proposed by D. Bernoulli. 
In order to investigate further this problem,
we consider the variation
of ${\cal F}$ induced by the variation
$\delta {\bf r}(s)$ of the position of the string. 
Since the filament is closed, there is no boundary 
terms and:
\begin{eqnarray}
\delta {\cal F}=
\int ds\; \delta {\bf r}.\partial_s{\bf A}
\label{e:variation_F}
\end{eqnarray}
where
\begin{eqnarray}
{\bf A}=
\left({C \over 2} \kappa^2 -\mu\right) {\bf t}
+C\partial_s\kappa \, {\bf n}
+C\kappa \tau \, {\bf b}\, .
\label{e:A}
\end{eqnarray}
Throughout the paper, we denote
the derivatives with respect to $s$ as $\partial_s$.
At equilibrium, the energy variation vanishes:
$\delta{\cal F}=0$, leading to the nonlinear
differential system $\partial_s{\bf A}=0$,
i.e. ${\bf A}$ is constant, and\cite{Doi}
\begin{eqnarray}
{C \over 2} \kappa^2 -\mu &=& {\bf A .t} \, ,
\label{e:var_E_1}
\\
-C\kappa \tau &=& {\bf A.b}\, .
\label{e:var_E_3}
\end{eqnarray}
The constant vector ${\bf A}$ 
represents the internal 
forces in the string \cite{Doi},
$\tau$ is the torsion, ${\bf t}=\partial_s{\bf r}$
is the tangent vector of the curve, and
$({\bf t, n,b})$ is the usual Frenet frame.
We have not written down the
projection of ${\bf A}$ on ${\bf n}$
because it is redundant. Indeed, it
and can be obtained from a derivation
of (\ref{e:var_E_1}) with respect to $s$.

But $\delta{\cal F}$
does not vanish only at the global minimum
(i.e. at equilibrium), and
a number of other spurious  solutions
are found\footnote{The reader who is not familiar
with the calculus of variations may consider
similar but simpler statement: The fact the derivative
of a function $f(x)$ of a scalar $x$ 
vanishes does not necessarily mean that we have reached the global
minimum.}. 
Planar solutions  of (\ref{e:var_E_1},\ref{e:var_E_3}) were  
analyzed by Euler \cite{Euler}. The 3D solutions are listed
in Ref.\onlinecite{Langer}. Comparing their energies, 
one finds that the circle is the 
closed solution with the lowest energy.
It is also the only closed solution
which is stable \cite{Langer}. Thus, 
knots cannot be stable solutions of (\ref{e:var_E_1},\ref{e:var_E_3}).
Knots can nevertheless be stabilized in the
presence of interactions
between the different parts of the string. 

In the following, we use a hard core repulsion,
and the string is modelled
by means of a non-self-intersecting 
tube of diameter $w$. Such a hard core
repulsion implies that
the distance between different parts of the 
string is larger than $w$,
and the radius of curvature $1/\kappa$
is larger than $w/2$ (see e.g.
Ref. \onlinecite{O'Hara2003}).
In the previous paragraph, we concluded that
knots could not be stabilized in the absence of interactions.
In the case of hard core repulsion,
such a statement means that
contact points must be present.

But the non-crossing condition at contact points involves 
interactions between distant parts of the
string (i.e. parts with different values of $s$).  
These interactions can be included
in the energy (\ref{e:E}) by means of an additional term
which includes a nonlocal interaction potential. 
In the variational formulation, such a term leads to additional
nonlinear integro-differential contributions to
(\ref{e:var_E_1},\ref{e:var_E_3}), the consequences of which 
are difficult to analyze. Despite the difficulty, some
mathematical informations about the solutions --
such as  their existence--
were obtained from this approach \cite{vonderMosel1998}.

We here consider a different
approach based on a combined analysis of
the knot topology and 
of the geometry in the limit of vanishing string
width $w\rightarrow 0$. Our approach does not
systematically provide the equilibrium configuration
and its energy, but it
allows one to obtain important information about the 
equilibrium energy (such as lower and upper bounds),
and about the geometry of the equilibrium 
configuration (such as the absence of knot localization
and the presence of braid localization).

\section{Lower bound for the energy}
\label{s:lower_bound}

\begin{figure}
\centerline{
\hbox{\psfig{figure=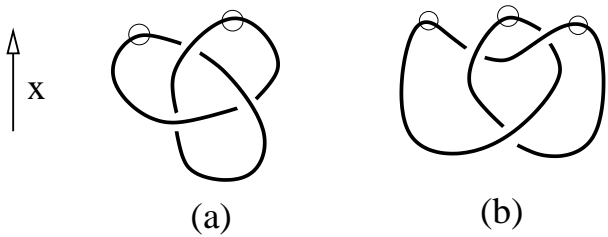,width=8cm,angle=0}}  }
\caption{The trefoil knot can be represented in
different ways in 3D space such as in (a) and (b).
For each configuration, there is a given
number of maxima along the $x$ axis: 2 in (a),
and 3 in (b). The minimum number of maxima among all
configurations of the knot  is the bridge number $n$.
On the figure, we have drawn the trefoil knot, for which $n=2$.
}
\label{fig:bridge_number}
\end{figure}

\begin{figure}
\centerline{
\hbox{\psfig{figure=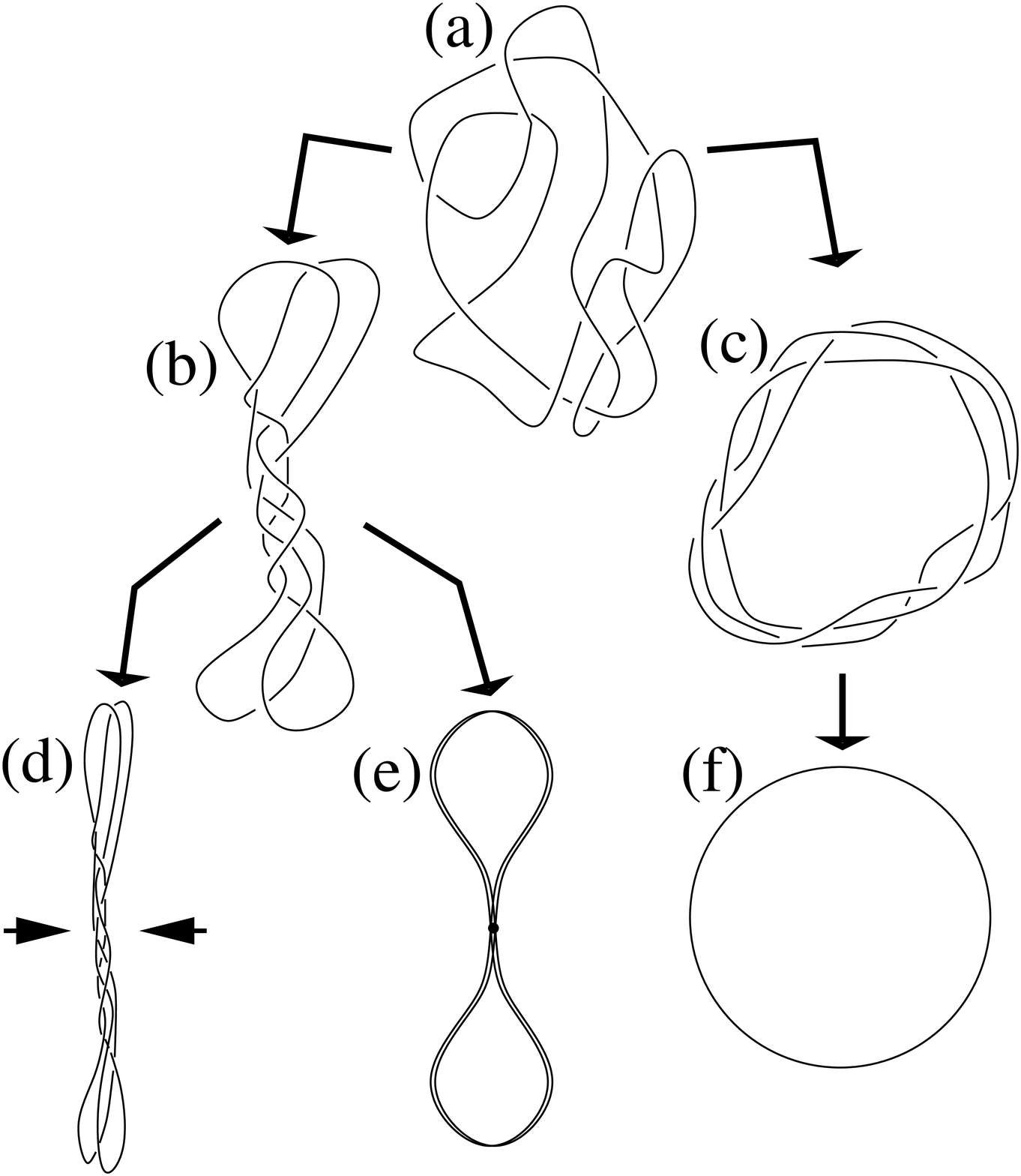,width=7cm,angle=0}}  }
\caption{ Knots in
ordinary 3D space.
Any knot (a) can be deformed to a braid and loop
configuration (b) with $n$ maxima,
or to a closed braid configuration (c) with
$i$ strands (here $n=2$ and $i=3$).
The lateral shrinking of (b) leads to the configuration
(d) with the lowest total curvature $\bar\kappa=2\pi n$.
When the central braid of (b) is shrinked to a point,
and the loops are solutions of Eqs.(\ref{e:var_E_1},\ref{e:var_E_3}),
we obtain the Point Braid and Loops configuration (e).
In (f), the closed braid is laterally shrinked, and has a circular
shape.
}
\label{fig:knot_transformations}
\end{figure}

We start with a result found by
J.W. Milnor \cite{Milnor1950}:
for any given knot $K$,
\begin{eqnarray}
2 \pi n \leq \bar\kappa  \, ,
\label{e:bridge_k}
\end{eqnarray}
where 
\begin{eqnarray}
\bar\kappa=\int ds \kappa
\label{e:total_curv}
\end{eqnarray}
is a dimensionless quantity called the
total curvature\cite{O'Hara2003},
and $n$ is the  bridge number of the knot $K$.
To define $n$, let us consider a given direction
in space $\hat {\bf x}$. As shown in Fig.
\ref{fig:bridge_number}, any given configuration 
${\bf r}(s)$ of $K$
in space has a well defined 
number of maxima along $\hat {\bf x}$. The minimum 
number of maxima among all configurations of $K$ 
is $n$. 
Deforming the knot so as to place all maxima
and all minima in two parallel planes of fixed abscissa 
along $\hat {\bf x}$, we obtain the 
linear braid configuration of figure \ref{fig:knot_transformations}b.
The minimum number of loops at the top of the braid is $n$.
For an unknotted closed string (usually called the unknot), 
$n=1$, and for
any other knot $n\geq 2$. For example, $n=2$ for most DNA knots
\cite{Wasserman1986}.

We define the normalized energy
\begin{eqnarray}
\epsilon = {{\cal E} {\cal L} \over C }
\label{e:epsilon}
\end{eqnarray}
which depends on the knot shape, but is independent of
the knot size and bending rigidity. Hence,
$\epsilon$ only depends on $w/{\cal L}$
and on the type of knot. Using the Schwarz inequality:
\begin{eqnarray}
\left(\int ds\, \kappa^2\right)\left(\int ds\right) 
\geq \left(\int ds\, \kappa \right)^2= \bar\kappa^2
\end{eqnarray}
and (\ref{e:bridge_k}), 
we find:
\begin{eqnarray}
2 \pi^2 n^2 \leq {\bar\kappa}^2/2 \leq \epsilon\, .
\label{e:epsilon_sup_n2}
\end{eqnarray}
This generalizes a result of Ref.\onlinecite{kamien},
which reads in our notations: $\epsilon\geq 8 \pi^2$,
and which can be obtained from (\ref{e:epsilon_sup_n2})
with the additional knotting condition $n\geq 2$.

Note that (\ref{e:bridge_k}) is sharp:
for any knot, it becomes an equality for the laterally
shrinked configuration of Fig.2d. Nevertheless,
(\ref{e:epsilon_sup_n2}) is not necessarily sharp:
it only
provides a lower bound for the energy.
Since it is a lower bound for all configurations,
it is also a lower bound for the 
equilibrium energy.
This lower bound depends on the knot type via $n$ 
and but does not depend on the filament width $w$.

\section{Asymptotics of thin strings and braid localization}

\subsection{Variation of the energy with $w$}
\label{s:w_var}
The energy ${\cal E}^*_w$ of the equilibrium
configuration
of a knot $K$ is a monotonically increasing function of $w$.
The proof of this statement
is reported in Appendix A.
We may therefore write:
\begin{eqnarray}
\partial_w{\cal E}^*_w\geq 0
\end{eqnarray}
Hence, if we consider the equilibrium shapes ${\cal E}^*_w$
of a given knot $K$ as a function of $w$, the lowest
value of ${\cal E}^*_w$ will be reached in the limit
$w\rightarrow 0$.
On the opposite, the 
highest value of ${\cal E}^*_w$ will be reached
by the tight knot\cite{Degennes}, 
also called ideal knot\cite{Staziak}, which
is the configuration with the highest possible
value $w_{id}$ of $w$ authorized by the hard core repulsion.
This can be summarized as:
\begin{eqnarray}
{\cal E}^*\leq {\cal E}^*_w
\leq {\cal E}^*_{w_{id}} .
\label{e:0_to_id}
\end{eqnarray}
where ${\cal E}^*=\lim_{w\rightarrow 0}{\cal E}^*_w$.

Here, we will not analyze the full dependence
of ${\cal E}^*_w$ with respect to $w$,
but we rather focus of the limit $w\rightarrow 0$.
The reason of our focus on this limit
is self-consistency. Indeed, the expression
of the energy ${\cal E}$, defined in
Eq.(\ref{e:E}), is valid in the limit (\ref{e:small_deformations}).
Integrating  (\ref{e:small_deformations}) along the knot,
we find $w\bar\kappa\ll{\cal L}$. Then, using
(\ref{e:bridge_k}),  we obtain
\begin{eqnarray}
{w \over {\cal L}} \ll {1 \over 2\pi n} \, .
\label{e:self_consistency_E}
\end{eqnarray}
Since we consider a given knot (i.e. fixed $n$)
with a fixed length (i.e. fixed ${\cal L}$), 
the limit $w\rightarrow 0$ is required 
for the energy ${\cal E}_w$ of a knot 
to be well-defined.

\subsection{Configurations in the limit $w\rightarrow 0$}
\label{s:w_to_0}
We  now
analyze the knot configurations 
in the limit $w\rightarrow 0$.
They are obtained from a procedure in 2 steps.
(i) Identification a possible
structure $\{{\cal S}\}$ of the knot 
when $w\rightarrow 0$.
(ii) Variational
approach on the structure $\{{\cal S}\}$.

\subsubsection{Structure}

In  step (i), we take the
limit $w\rightarrow 0$ which may lead
to ``wild knots", exhibiting knot accumulations 
or singularities. We shall determine which
types of accumulations or singularities are allowed, 
and which ones are forbidden.

\begin{figure}
\centerline{
\hbox{\psfig{figure=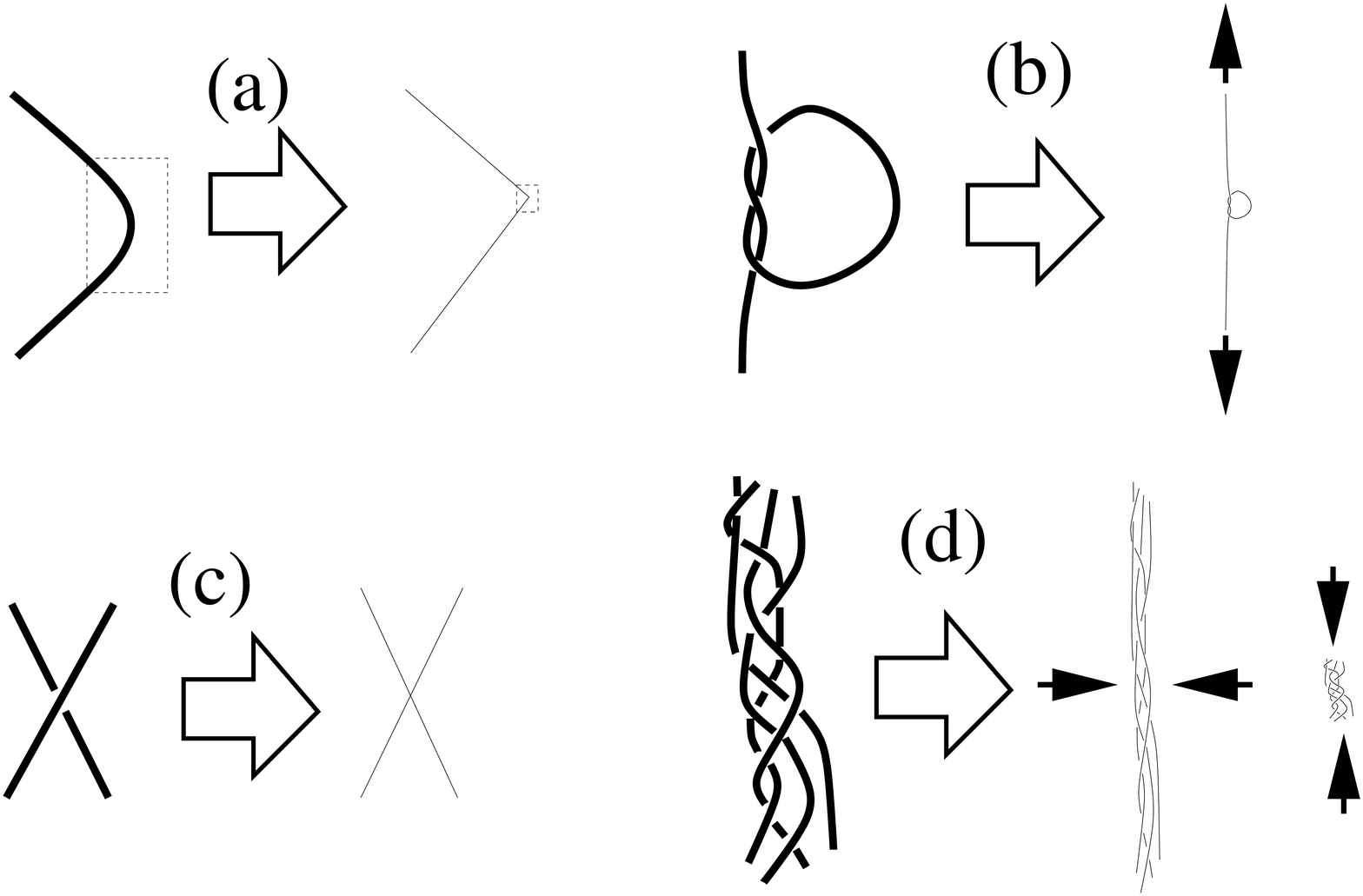,width=7cm,angle=0}}  }
\caption{
In the limit where the string width $w$ vanishes,
various singularities or accumulations may occur.
The thickness of the lines are used to represent various
filament widths $w$.
(a) angle. The dashed line defines a box whose
size is shrinked to zero to obtain localization.
The filament length in the box is ${\cal L}_{loc}$.
see text; (b) knot localization; (c) multiple point;
(d) braid localization, leading to a line-braid,
or a point braid. Angles and knot localization
are forbidden because they lead to divergence of the
energy ${\cal E}$. 
}
\label{fig:accum}
\end{figure}

The 4 possible types of singularities or accumulations
are given on Fig.\ref{fig:accum}.
Angular points  --as in Fig.\ref{fig:accum}a, and
knot localization --as in Fig.\ref{fig:accum}b,
are forbidden because they
lead to an infinite energy.
A rigorous proof of this statement
is given in Appendix \ref{a:angle_knot_localization}.
Here, we only provide an intuitive explanation:
angles and knot localization
may be obtained by decreasing 
the length ${\cal L}_{loc}$ of a part
of a curve to zero, keeping its 
shape fixed. An example of such a 
shrinkage is shown on Fig. \ref{fig:accum},
where the size of the dashed box containing the angle decreases.
Using Eq.(\ref{e:epsilon}), the energy of this part
behaves as ${\cal E}_{loc}=C\epsilon/{\cal L}_{loc}$
where $\epsilon$ depends on the shape but not
on the scale. Since ${\cal L}_{loc}\rightarrow 0$, one has
${\cal E}_{loc}\rightarrow\infty$.

The two types of accumulations which are allowed
when $w\rightarrow 0$ are multiple points 
--as in Fig. \ref{fig:accum}c, or
braid localization --as in Fig.\ref{fig:accum}d, because
they do not lead to a divergence of ${\cal E}$. 

Braid localization is defined as the
lateral shrinking of a braid, with all strings
in the braid tending to the same curve.
We define a line-braid of multiplicity $m\geq 1$ as 
the result of the lateral shrinking of a braid
of $m$ strings. In the following, a single string
will be considered as a line-braid with 
multiplicity $m=1$. Since all strings in the braid
tend to the same curve, the curvature energy of the line-braid
reads $m C/2\int ds\, \kappa^2$.

In some cases, we might then also shrink the length
of the braid to zero to obtain a point-braid,
as in Fig. \ref{fig:accum}d.
The order of the limits is important: firstly
lateral shrinking, and secondly length shrinking,
so that the curvature energy of the point-braid vanishes.
Since angle localization is forbidden, 
strings or line-braids must emerge tangentially
along the same axis from a point-braid.

\begin{figure}
\centerline{
\hbox{\psfig{figure=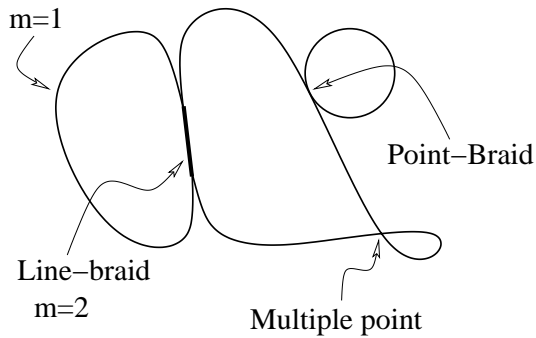,width=7cm,angle=0}}  }
\caption{Shape of a knot in the limit $w\rightarrow 0$:
an ensemble of line-braids connected by 
multiple points or point-braids. The drawing is 2D but 
we consider 3D configurations throughout the manuscript.
}
\label{fig:structure}
\end{figure}

By shrinking braids in a special way
with a given knot, we obtain
a structure $\{{\cal S}\}$ of 
line-braids connected
by multiple points or point-braids. 
An example of such a structure is 
depicted on Fig.\ref{fig:structure}.

At this point, we shall  put some 
emphasis on a central result
of the present section: in the limit $w\rightarrow 0$,
knot localization is forbidden and braid localization
is expected. We shall also stress on the fact that
we have not proved that braid localization {\it will} occur,
but we have shown that it {\it can} occur when $w\rightarrow 0$.

\subsubsection{Variational analysis of a structure}
\label{s:variational}

In this section, we derive the conditions which
must be satisfied at equilibrium from a variational
approach. 
We consider a given structure $\{{\cal S}\}$,
with $N$ line-braids. Its energy reads:
\begin{eqnarray}
{\cal E}_{\{{\cal S}\}}=\sum_{p=1}^N {m_p C\over 2} \int_p ds \kappa^2
\label{e:E_P}
\end{eqnarray}
where the index $p$ in the integral means that the integration
is performed over the $p$th line-braid only,
and $m_p$ is the multiplicity of the $p$th line-braid.
As in section \ref{s:model}, the total length
is fixed by means of a Lagrange multiplier $\mu$,
and we have to minimize:
\begin{eqnarray}
{\cal F}_{\{{\cal S}\}} =\sum_{p=1}^N \left[{m_p C\over 2} \int_p ds \kappa^2
+\mu m_p \int_p ds \right]
\label{e:F_P}
\end{eqnarray}
The variation of ${\cal F}_{\{{\cal S}\}}$ resulting from a
variation $\delta{\bf r}(s)$ of the structure
$\{{\cal S}\}$ reads:
\begin{eqnarray}
\delta {\cal F}_{\{{\cal S}\}}=\sum_{p=1}^N && \left\{
\int_p ds \; \delta {\bf r}.\partial_s{\bf A}_p \right.
\nonumber \\
&&+ \left. \left[-\delta {\bf r}.{\bf A}_p
+Cm_p\kappa \,{\bf n}.\partial_s\delta {\bf r}
\right]_p\right\}
\label{e:variation_F_P}
\end{eqnarray}
where $[Y]_p$ indicates the difference between $Y$ 
at the end of the $p$th line braid, and $Y$ at its beginning.
Moreover,
\begin{eqnarray}
{\bf A}_p=m_p\left[ 
\left({C \over 2} \kappa^2 -\mu\right) {\bf t}
+C\partial_s\kappa \, {\bf n}
+C\kappa \tau \, {\bf b} \right] \, .
\label{e:A_P}
\end{eqnarray}
At equilibrium, one has $\delta {\cal F}=0$, which leads to
$\partial_s{\bf A}_p=0$, so that
${\bf A}_p$ is a constant vector.
Projecting ${\bf A}_p$ on ${\bf t}$ and ${\bf b}$, we find:
\begin{eqnarray}
m_p\left[{C \over 2} \kappa^2 -\mu\right] &=& {\bf A}_p .{\bf t} \, ,
\label{e:var_E_1_P}
\\
-m_pC\kappa \tau &=& {\bf A}_p.{\bf b}\, .
\label{e:var_E_3_P}
\end{eqnarray}
At equilibrium, line-braids therefore obey  differential
equations similar to that 
of single strings  (\ref{e:var_E_1}-\ref{e:var_E_3}).

Let us now discuss the boundary conditions
at the contact points (multiple points
or point-braids) between the line-braids.
We use the index $B$ to list the points
at which the line-braids are connected.
At a point $B$, several strings 
related to different line-braids and point-braids
may cross. Two types of constraints may force
the strings which cross at $B$ to be tangent to each other.
(i) Since angle localization is forbidden,
each string enters and exits at the point $B$ 
with the same tangent vector. (ii) Each string is also
tangent to some other strings because it belongs
to a line-braid or a point-braid. The combination
of these two constraints forces a bunch of line-braids
ending at $B$
to be tangent to each other at the contact point.
There might be several bunches of line-braids at the point
$B$. These bunches can rotate freely
from each other, but all braids inside the
bunch are tangent to each other at the point $B$.
Let $q_B$ be the index which lists the bunches
at the point $B$.
The boundary contribution of the variation (\ref{e:variation_F_P})
may then be re-written as:
\begin{eqnarray}
&& \sum_{p=1}^N[-\delta {\bf r}.{\bf A}_p
+Cm_p\kappa \,{\bf n}.\partial_s\delta {\bf r}]_p
\nonumber \\
&=& -\sum_B  \delta{\bf r}_B.\sum_{p\rightarrow B}{\bf A}_p
\nonumber \\
&& +C\sum_B \sum_{q_B} \partial_s\delta {\bf r}_{q_B}. 
 \sum_{p\in q_B}
m_p\kappa \,{\bf n}\, ,
\label{e:grouping_bc}
\end{eqnarray}
where $p\rightarrow B$ indicate that we perform
the sum over all parts connected to $B$.
Moreover, ${p \in p_B}$ indicates that we perform
the sum over all line-braids and point-braids
belonging to the bunch $q_B$.
For definiteness, each braid should be oriented,
so that each term in the sums over $q\rightarrow B$ and $p\in q_B$
contains a sign $\pm$ \footnote{
As expected, the orientation does not affect the physics.
The reader interested in this point is invited to check the 
invariance of the results with respect to
the variable change $s\rightarrow -s$.}.
The quantities  $\delta{\bf r}_B$, and $ \partial_s\delta {\bf r}_{q_B}$
respectively account for infinitesimal translation of the
point $B$ and rotation of the bunch $q_B$ around the point $B$.
At equilibrium, one must have $\delta {\cal F}_{\{{\cal S}\}}=0$
for all perturbations, so that from Eqs.(\ref{e:variation_F_P})
and (\ref{e:grouping_bc}):
\begin{eqnarray}
\sum_{p\rightarrow B}{\bf A}_p &=& 0\, ,
\label{e:BC_P_A}
\\
\sum_{p \in q_B} m_p \kappa_p {\bf n}_p &=& 0\, .
\label{e:BC_P_Q}
\end{eqnarray}
The differential system (\ref{e:var_E_1_P}-\ref{e:var_E_3_P}),
together with the boundary conditions (\ref{e:BC_P_A}-\ref{e:BC_P_Q})
is well posed and should be solved in order to
determine the configurations of a given
structure $\{{\cal S}\}$ for which the
variation of the energy vanishes, i.e. $\delta {\cal F}_{\{{\cal S}\}}=0$.
In the following, we will denote the
solutions which obey the equation $\delta{\cal F}_{\{{\cal S}\}}=0$
with the index $^\circ$. For example, the
resulting energy of a structure
${\{{\cal S}\}}$ will be written as ${\cal E}^\circ_{\{{\cal S}\}}$.

As in the case of the Euler-Bernoulli elastica,
which was discussed in section \ref{s:interactions},
the class of configurations which obey the
equation $\delta {\cal F}_{\{{\cal S}\}}=0$
contains stable and unstable solutions.
Our goal here is not to analyze this class of configurations
in details. We will rather look for some specific 
configurations belonging to it, which will provide us with
upper bounds for the equilibrium energy.

We shall now present two relations which 
apply to this class of configurations. 
The first one provides the general expression
of the Lagrange multiplier $\mu^\circ_{\{{\cal S}\}}$
for a given structure $\{{\cal S}\}$:
\begin{eqnarray}
\mu^\circ_{\{{\cal S}\}}
={{\cal E}^\circ_{\{{\cal S}\}} \over {\cal L}}
= \epsilon^\circ_{\{{\cal S}\}}{C \over{\cal L}}^2\, ,
\label{e:mu_closed}
\end{eqnarray}
The demonstration of this relation
is reported in Appendix \ref{a:variational}.
The relation (\ref{e:tension_energy})
allows one to interpret $\mu^\circ$
as the average curvature energy density
(per unit length of string).

Another relation, which 
applies to a restricted class of structures,
will be very usefully in the following.
Indeed, in some cases, the normalized energy
of the structure can be expressed by means
of the normalized energies its line-braids:
\begin{eqnarray}
\epsilon^\circ_{\{{\cal S}\}}= \left[\sum_{p=1}^Nm_{p} 
\epsilon_p^{\circ 1/2}\right]^2 \, .
\label{e:epsilon*_parts}
\end{eqnarray}
and the length of the $p$th line-braid reads
\begin{eqnarray}
{\cal L}_p^\circ={\cal L} \left(
{\epsilon_p^\circ \over \epsilon^\circ_{\{{\cal S}\}} }
\right)^{1/2}\, .
\label{e:L_P_parts}
\end{eqnarray}
The situations where these formulas applies are: 
(i) all line-braids are arcs
of circles, (ii) all line-braids are closed loops,
and (iii) all line-braids have the same energy and the same length
(this includes the case of line-braids with identical shapes).
The derivation of this relation, as well
as the general expression of $\epsilon^\circ_{\{{\cal S}\}}$
are given in Appendix \ref{a:variational}.

\section{Upper bounds for the energy}

\subsection{Bridge number and braid index}
Let us consider a first example of configuration
belonging to the above-mentioned class.
We have seen in section \ref{s:lower_bound} that
any knot can be deformed in order to obtain
a configuration similar to that of Fig.\ref{fig:knot_transformations}b
with $n$ maxima. In the limit $w\rightarrow 0$, 
this configuration
can be deformed in such a way
to obtain the Point Braid and Loops (PBL) configuration 
of Fig.2e, which defines the structure $\{{\cal S}\}$.
Each loop has the same boundary conditions:
the curve begins and ends at the same point,
the initial, and final tangent vectors being opposite.
The detailed calculation of the minimum energy
of one loop is reported in Appendix
\ref{a:energy_calc}. We find  $\epsilon_{loop}^\circ\approx 18.19$.
Since the structure $\{{\cal S}\}$ is an ensemble
of identical loops, formulas 
(\ref{e:epsilon*_parts},\ref{e:L_P_parts}) apply.
Therefore, all loops have the same length and:  
\begin{eqnarray}
\epsilon^\circ_{PBL}=4n^2\epsilon_{loop}^\circ.
\label{e:epsilon*_SBL}
\end{eqnarray} 
This expression provides a first upper bound for the 
equilibrium energy.

A second upper bound
is found using Alexander's theorem \cite{alexander},
which stipulates that any knot can be 
transformed into a closed braid, as shown on
Fig.\ref{fig:knot_transformations}c.
The smallest possible number of 
strings in the closed braid is  the
braid index $i$\cite{Kauffman}, and $i\geq n$. In the limit $w\rightarrow 0$,
we may laterally shrink the closed braid with $i$ strings,
and we obtain a closed line-braid
of multiplicity $i$.
As mentioned in section \ref{s:interactions}, 
the equilibrium energy for a closed line
is reached by the circular configuration,
for which $\epsilon_0^\circ=2\pi^2$.
The value of $\epsilon^\circ_{CB}$ 
for a closed line-braid is once
again obtained from (\ref{e:epsilon*_parts}):
\begin{eqnarray}
\epsilon_{CB}^\circ=i^2\epsilon_0^\circ=2\pi^2i^2 \, .
\label{e:epsilon*_CB}
\end{eqnarray}

We see that for small $i$, i.e. when 
$n\leq i\leq \alpha n$, with
$\alpha=(2\epsilon_{loop})^{1/2}/\pi\approx 1.92$,
the CB configuration
has a lower energy than the PBL configuration. On the opposite,
for large $i$, i.e. when $i\geq \alpha n$,
the PBL configuration has the lowest energy.
Combining the lower bound (\ref{e:epsilon_sup_n2}) 
and the upper
bounds (\ref{e:epsilon*_SBL},\ref{e:epsilon*_CB})
we find that the normalized equilibrium energy
obeys:
\begin{eqnarray}
2 \pi^2 n^2 \leq \epsilon^*
\leq 2 \pi^2
\min[ \alpha^2 n^2; i^2] \, .
\label{e:epsilon_ineq_0} 
\end{eqnarray}
Using (\ref{e:epsilon}), we finally have:
\begin{eqnarray}
2 \pi^2 n^2{C\over {\cal L}} \leq {\cal E}^* 
\leq 2 \pi^2
\min[ \alpha^2 n^2; i^2]{C \over {\cal L}}
\label{e:epsilon_ineq}
\end{eqnarray}
These inequalities are a central statement 
of the present paper.
Let us now present two results which directly follow
from them.

\subsection{Scaling of the equilibrium energy}

First, the inequalities (\ref{e:epsilon_ineq}) allows us to reach
a general conclusion: the equilibrium energy  exhibits
upper and lower bounds both proportional to $n^2{C/{\cal L}}$.
We shall write this result as:
\begin{eqnarray}
{\cal E}^* \sim n^2{C\over {\cal L}}, 
\end{eqnarray}
Although this relation does not provide the exact
value of ${\cal E}^*$, it is a strong indication
of the behavior of ${\cal E}^*$ as a function of knot
complexity. For example, we may conclude that 
${\cal E}^*$  can be large only for knots
having a large value of $n$.

\subsection{The $n=i$ knot family}
The second consequence of (\ref{e:epsilon_ineq})
points out a specific class of knots. Indeed, when $n=i$,
(\ref{e:epsilon_ineq}) becomes an equality, and 
our problem
is readily solved: we  have
\begin{eqnarray}
{\cal E}^*=2\pi^2i^2{ C\over {\cal L}}
\end{eqnarray}
and the configuration is a circular line-braid
of multiplicity $i$. 
The relation $n=i$ defines a knot family
which contains the torus knots
(knots obtained by wrapping
a string on the surface of a torus
without crossing).
Using available tables \cite{knot-atlas} we have  analyzed
prime knots of crossing number $n_c<11$.
The crossing number $n_c$ is defined as the smallest number
of crossings in the planar projections of a knot.
We find that the $n=i$ knot family contains
$\approx 20\%$ of these knots.

\section{Monte Carlo Simulations}
\label{s:MC}
In order to gain more insights about the
configuration of stiff knots, we have performed
Monte Carlo (MC) simulations with
a closed chain of $N=150$ beads 
separated by segments of fixed 
length --equal to 1. The length-preserving
elementary motion of the 
chain is implemented via the rotation 
(with angle $\pm \pi/100$) of
one bead around the axis which
runs through its neighbors. 
We use the
Metropolis algorithm, with the energy:
\begin{eqnarray}
{\cal E}_d=C\sum_{n=1}^N(1-{\bf u_n. u_{n+1}})\, ,
\end{eqnarray}
where ${\bf u_n}$ is a unit vector along
the $n$th  segment. The closure of the chain imposes:
\begin{eqnarray}
\sum_{n=1}^N {\bf u}_n=0 \, ,
\\
{\bf u_{N+1}}={\bf u}_1.
\end{eqnarray}
At low temperatures, the chain length
${\cal L}=N$ is much smaller
than the persistence length $L_p=C/k_BT$.
Then, the curve is smooth, 
and ${\cal E}_d\rightarrow {\cal E}$.
Non-crossing conditions are imposed with 
spheres of excluded volumes around
each bead: we forbid
beads to get closer than $1/\sqrt{2}$. 
This leads to an excluded volume tube
with a non-constant width $w$,
varying between $1$ and $\sqrt{2}$.
We will analyze knots
with $n=2$ only, so that the condition
(\ref{e:self_consistency_E}) is verified:
$w/{\cal L}\sim 10^{-2}\ll 1/2\pi n\sim 10^{-1}$.

We use a simulated annealing method
with  a power law decrease of the temperature
up to the low temperature regime.
Repeated simulations with the same knot provide
us with the ground state and sometimes also
with metastable states.

\begin{figure}
\centerline{
\hbox{\psfig{figure=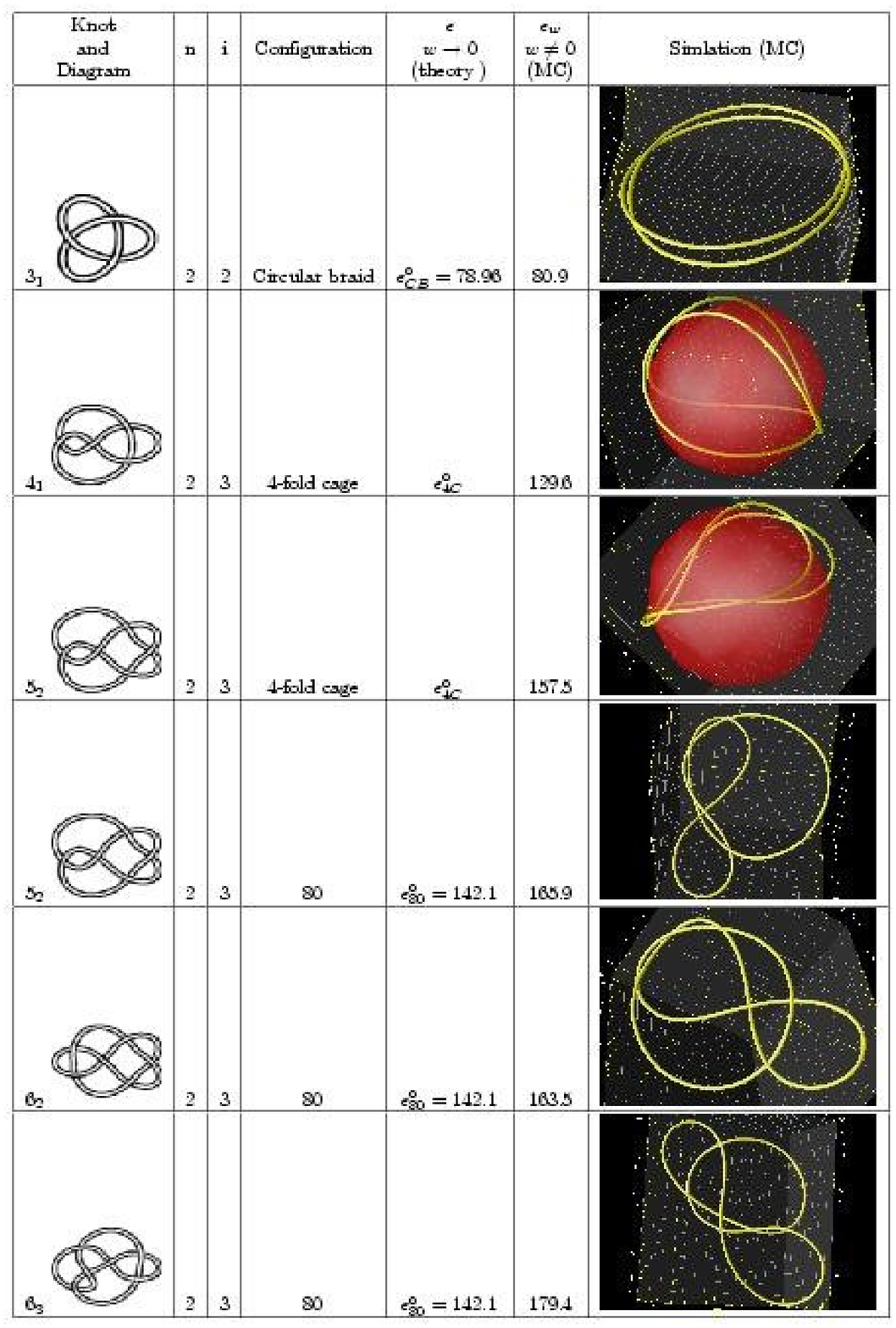,width=8cm,angle=0}} 
}
\caption{
Table of results for the MC simulations of closed
stiff knots. Three different configurations are found:
(i) the circular braid for the $3_1$ knot;
(ii) the 4-fold cage for the $4_1$ and $5_2$ knots
(the sphere is a guide for the eye, indicating that this
configuration is approximately wrapped around a sphere);
(iii) the 80 for the $5_2$, $6_2$ and $6_3$.
}
\label{fig:MC}
\end{figure}

\begin{figure}
\centerline{
\hbox{\psfig{figure=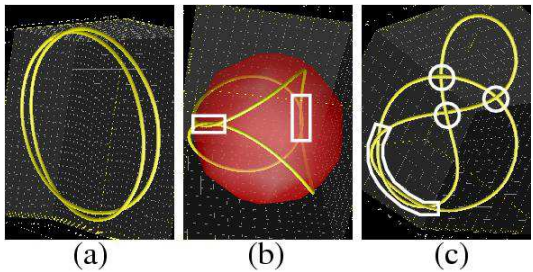,width=7cm,angle=0}}  }
\caption{
(a) The $3_1$ provides an example of line-braid of
multiplicity 2. (b) The two contact zones in the 4-fold cage
configuration are point-braids, indicated by
white boxes. (c) An 80 configuration,
exhibiting one point-braid and 3 double points,
indicated by small circles.
}
\label{fig:MC_braids}
\end{figure}

A list of the knots studied in the simulations,
together with the obtained equilibrium configurations,
is presented in Fig.\ref{fig:MC}.
A knot is usually denoted
with its crossing number $n_c$, and with
an index which indicates its order in the
standard list of knots for a given value of $n_c$.
For the trefoil knot --denoted as $3_1$,
which is a torus knot with
$n=i=2$, the expected circular line-braid configuration
is found, as shown  on Fig.\ref{fig:MC}.
Fig.\ref{fig:MC} also shows two other configurations:
the $4_1$ (figure 8 knot), and
$5_2$ knots lead to a configuration
which will be denoted as the 4-fold cage configuration
in the following.
The $5_2$, $6_2$, $6_3$
knots lead to a configuration
which we call the 80 configuration
(because it is composed of an 8
and a circle).
The three different shapes are summarized
on Fig.\ref{fig:MC_braids}. We observe that braid
localization is present in all of them:
line-braid for the $n=i$ configuration, 
and point braids for the two other configurations.

\begin{figure}
\centerline{
\hbox{\psfig{figure=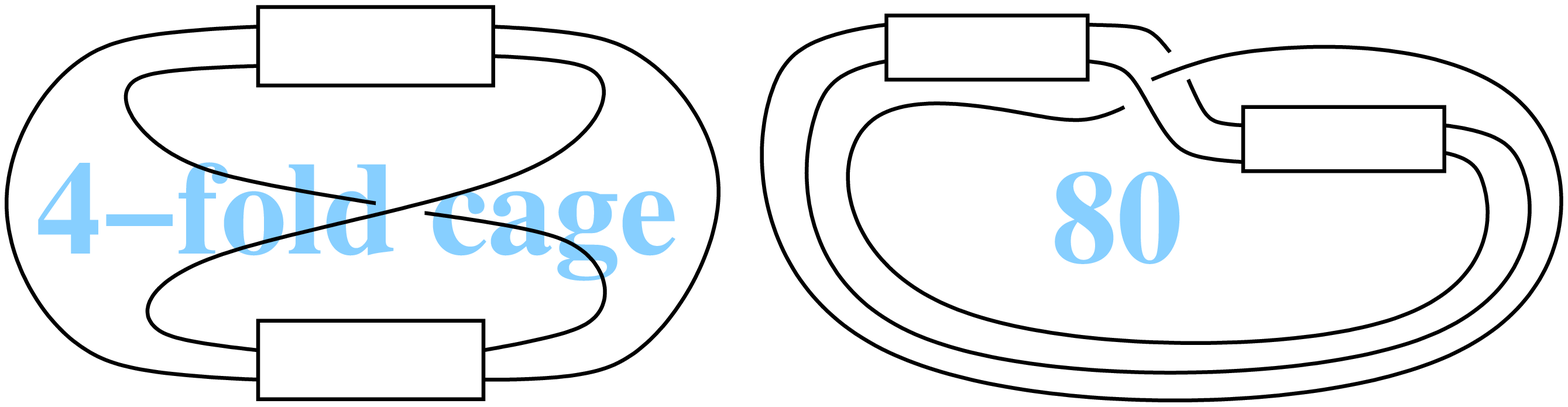,width=5cm,angle=0}}  }
\caption{
Knot families leading to 4-fold cage
and 80. The boxes account for twists
with an arbitrary number of turns.
}
\label{fig:conjectured_families}
\end{figure}

In Fig.\ref{fig:conjectured_families}, 
we have shown the conjectured
knot families which
lead to the 4-fold cage and the 80. 
These conjectured families are also reported
on Fig.\ref{fig:MC}.
From knot tables \cite{knot-atlas},
the sum of the 3 families
$n=i$, 4-fold cage, and $80$ represent
most of simple knots:
all prime knots with $n_c<7$,
and $75\%$ of those 
with $n_c<9$.

We have analyzed in details
the energy of the 80 configuration. 
It is calculated from the
procedure of section\ref{s:w_to_0}. 
The structure of the 80 is composed of two
parts: the 8, and the circle 0. The energy of the
8 curve is calculated in Appendix \ref{a:energy_calc}:
 $\epsilon_8^\circ\approx 55.93$. The 0 is a circle,
and its energy is $\epsilon_0^\circ=2\pi^2$.
Since the 0 and the 8 are independent planar solutions,
we may assume vanishing forces at their contact points,
and analyze the structure of the 80 as a sum
of two closed solutions (the 0 and the 8).
Using Eq.(\ref{e:epsilon*_parts}), we may then evaluate
the equilibrium energy of the 80 structure as:
\begin{eqnarray}
\epsilon_{80}^\circ=(\epsilon_8^{\circ 1/2}
+\epsilon_0^{\circ 1/2})^{2}\approx 142.1.
\end{eqnarray}
As expected from section \ref{s:w_var}, the energy
in the MC simulations (given on Fig.\ref{fig:MC}), 
corresponding to a finite (but small) value of the
width $w$, is slightly larger than the theoretical result
corresponding to the limit $w\rightarrow 0$.

From the MC simulations,
a normalized energy equal to $129.6$ is obtained
for the 4-fold cage. Since $\partial_w{\cal E}^*_w>0$,
this means that in the limit $w\rightarrow 0$,
$\epsilon^*_{4C}<129.6$. The value of $\epsilon^*_{4C}$
is probably quite close to its numerical upper bound $129.6$,
but we do not know its exact value.
We only know from (\ref{e:epsilon_ineq_0}) that,
since $n=2$, one has $\epsilon^*_{4C}>8\pi^2$.
Finally, the different configurations relevant for the
simulations fall into the hierarchy:
\begin{eqnarray}
\epsilon^\circ_{CB}|_{i=2}<\epsilon^\circ_{4C}<\epsilon^\circ_{80}
<\epsilon^\circ_{CB}|_{i=3}
<\epsilon^\circ_{PBL}|_{n=2} \, .
\label{e:hierarchy}
\end{eqnarray}
As seen from Fig.\ref{fig:MC},
some knots may exhibit more than one configuration. 
For example, both the 4-fold cage and the 80
configurations where observed with the $5_2$ knot.
The hierarchy (\ref{e:hierarchy}) shows that
the 4-fold cage is the ground state, 
while the $80$ is metastable. This result
is also observed in the simulations:
although the energies of both configurations
are slightly larger than the asymptotic
value for $w\rightarrow 0$, 
their order is not affected by the
finiteness of the width $w$.

\section{Open knots}
\label{s:open_knots}

Let us now consider an open knot,
which exhibits two asymptotically straight open ends far from
the knot as in Fig.1. We may ask the same question
as for close knots: what are the configuration
and the energy at equilibrium?
We shall here show that our results on closed
knots can be extended to the case of 
open knots. 

\begin{figure}
\centerline{
\hbox{\psfig{figure=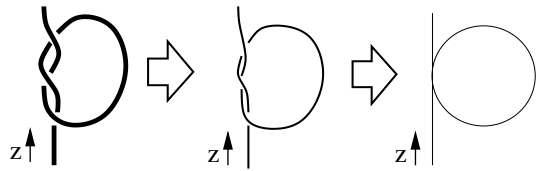,width=7cm,angle=0}}  }
\caption{
Schematics of the open trefoil stiff knot.
Open ends are along the $z$ axis.
From left to right, the filament width $w$ decreases.
}
\label{fig:open_trefoil}
\end{figure}

For an open knot, the total length
of the string diverges. It is therefore more suitable to use
the excess length ${\cal L}_e$, defined as the total length
minus the length of a straight string:
\begin{eqnarray}
{\cal L}_e=\int ds (1-{\bf t} \cdot {\bf \hat z})
=\int ds-\int dz
\label{e:excess_length}
\end{eqnarray}
where the integrals are performed along the
whole knot, and ${\bf \hat z}$ is the direction
of the open ends far from the knot (We assume that
both ends have the same direction).
The excess length ${\cal L}_e$ should replace ${\cal L}$
in the expression of the energy ${\cal F}$
defined in Eq.(\ref{e:F}).
The variations are performed in the same way,
and since the last term of (\ref{e:excess_length}) is constant,
its variation vanishes, and the obtained differential
system is the same as that for close knots.

Since open ends are straight,
we assume that $\kappa\rightarrow 0$, 
and $\partial_s\kappa\rightarrow 0$
far from the knot.
Then, from (\ref{e:A_P}) we have ${\bf A.t}\rightarrow \mu^\circ$,
${\bf A.n}\rightarrow 0$, and ${\bf A.b}\rightarrow 0$.
Since ${\bf A}$ describes internal forces in the string,
and ${\bf t}$ is the tangent vector,
this means that $\mu^\circ$ is the equilibrium tension far
from the knot.

For a given structure $\{{\cal S}\}$
of an open knot at equilibrium, we show 
in Appendix \ref{a:variational} that the Lagrange multiplier
$\mu^\circ$ is related to the curvature energy:
\begin{eqnarray}
\mu^\circ_{\{{\cal S}\}}
={{\cal E}^\circ_{\{{\cal S}\}} \over {\cal L}_e}
= \epsilon^\circ_{e\{{\cal S}\}}{C \over{\cal L}_e^2}\, ,
\label{e:tension_energy}   
\end{eqnarray}
where the normalized energy of an open knot is defined as:
\begin{eqnarray}
\epsilon^\circ_{e\{{\cal S}\}}={{\cal E} \over {\cal L}_e} \, .
\label{e:epsilon_e}
\end{eqnarray}
The general expression of 
$\epsilon^\circ_{e\{{\cal S}\}}$ is given in 
Appendix \ref{a:variational}.
Note that relation (\ref{e:tension_energy}) is also valid for
closed knots, as shown in section \ref{s:variational}.
Since the excess length
${\cal L}_e$ is fixed, Eq.(\ref{e:tension_energy}) 
also shows that the values of
$\mu^\circ$  are ordered
in the same way as the energies.

For open knots, the bridge number $n_1$ 
and the braid index $i_1$ are defined
in the same way as for closed knots.
If an open knot is obtained by opening
a closed knot with bridge number $n$
and braid index $i$, the bridge number
and braid index of the open knots are \cite{open-closed}:
\begin{eqnarray}
n_1=n-1\, ,
\\
i_1=i-1\, .
\end{eqnarray}
As an example, since $n=i=2$ for the trefoil knot,
one has $n_1=i_1=1$ for the open trefoil knot.

The PBL configuration of Fig.\ref{fig:knot_transformations}e
with open ends connected  to the knot via a straight line
tangent to the point braid, is used as a first upper bound.
A second upper bound is obtained with the circular braid
configuration of Fig.\ref{fig:knot_transformations}f
with open ends connected  to the knot via a straight line
tangent to the circular line-braid.
Finally,
(\ref{e:epsilon_ineq}) and (\ref{e:tension_energy}) are also valid
for open  knots when ${\cal L}$ is replaced by the
excess length ${\cal L}_e$,
so that  the equilibrium tension $\mu^*$ of open knots obeys
\begin{eqnarray}
2 \pi^2 n_1^2 {C\over  {\cal L}_e^2} \leq \mu^* \leq 2 \pi^2
\min[ \alpha n_1^2; i_1^2] {C\over {\cal L}_e^2} \, .
\label{e:mu_ineq}
\end{eqnarray}
We therefore conclude that the equilibrium tension
behaves as $n_1^2$, i.e. $\mu^*\sim n_1^2C/ {\cal L}_e^2$.
Furthermore, when $n_1=i_1$, the equilibrium shape of open knots
is a circle tangent to a point braid,
and  $\mu^*=2\pi^2i_1^2 C/{\cal L}_e^2$.

The simplest example of open knot is the
open trefoil knot ($3_1$), for which $n_1=i_1=1$.
As depicted in Fig.\ref{fig:open_trefoil}, 
our theory predicts  a circle tangent
to a straight string in the limit $w\rightarrow 0$.
This familiar shape is indeed easily obtained
with a nylon string, a metal string, or hair.
An example with a Silica nanowire is presented
in Fig.\ref{fig:phys_knots}.
The relation between tension and
excess length $\mu^*=2\pi^2 C/{\cal L}_e^2$ for
the open trefoil knot was in fact already used as an ansatz
in Ref.\cite{actin}, where 
it was experimentally checked 
and used to evaluate the bending
rigidity of actin. Here we show that it is 
an exact result in the limit 
$w\rightarrow 0$. Checking the $n^2$ dependence
of $\mu^*$ by varying the knot type
opens a novel and challenging line of investigations
for experiments.

\section{Discussion}

In the following, we shall make some remarks,
and briefly mention some open issues 
related to the present work.

\subsection{
On the limits $w\rightarrow 0$ and ${\cal L}\rightarrow\infty$}

Throughout the present paper, we have analyzed the limit $w\rightarrow 0$.
Nevertheless, in a given experimental situation, it is difficult
to perform a variation of the width of the filament.
Furthermore, we have assumed that
the bending rigidity modulus does not
vary with the width. But it often does,
as e.g. in continuum elasticity\cite{Doi}, where $C\sim w^{-4}$.
A more natural limit for experiments would be
to take ${\cal L}\rightarrow \infty$ with fixed $w$. Both limits
are equivalent. Indeed, the important
point is that $w/{\cal L}\rightarrow 0$, 
as seen e.g. in the inequality (\ref{e:self_consistency_E}).

\subsection{
Curvature energy of thick knots}

Two upper bounds for the energy of stiff knots
can be derived from the limit of finite  $w$.
First, a general upper bound directly 
follows from
excluded volume effects. Indeed, we have seen in
section \ref{s:interactions} that $\kappa\leq 2/w$.
Hence,
\begin{eqnarray}
{\cal E}={C \over 2}\int ds \kappa^2\leq 2C {{\cal L} \over w^2}\, .
\end{eqnarray}
which may be re-written in terms of
the dimensionless normalized energy:
\begin{eqnarray}
\epsilon= {{\cal E}C \over {\cal L}}\leq
2 \left({{\cal L}\over w}\right)^2\, .
\end{eqnarray}

Using (\ref{e:0_to_id}), one may also use
${\cal E}_{id}$ as an upper bound for
the energy of a given knot. We do not know the 
precise value of ${\cal E}_{id}$, which is
dictated by the geometry of the 
ideal knot.
Nevertheless,  a lower bound for ${\cal E}_{id}$ may be found
from  a recent conjecture based on
results for lattice knots\cite{Ernst}, which states that
\begin{eqnarray}
\bar\kappa_{id}\geq b \, n_c^{1/2}
\end{eqnarray}
where $b$ is a positive constant. 
Using the Schwarz inequality as in section
\ref{s:lower_bound}, we find that:
\begin{eqnarray}
{\cal E}_{id}\geq {b^2\over 2}n_c {C \over {\cal L}}
\end{eqnarray}
which may also be written as
$\epsilon_{id} \geq b^2n_c/2$.

\subsection{Experimental shapes for stiff knots}
\label{s:exp}

We have performed rudimentary experiments
in order to corroborate the results of our Monte Carlo simulations.
We do not look for quantitative measurements,
but we rather aim for a qualitative confirmation
of the theory and simulations.

The experiments are performed with a plastic tube of width
$w=1$cm, and length ${\cal L}=80$cm. To close the tube,
we have joined the two ends using a small stick
inserted in both ends. This closure allows
tangential matching as well as
free local rotation of one end with respect to
the other at the contact point. Therefore,
the tube cannot store twist, and the torsion
energy is neglected. These experiments are 
imperfect, and cannot be used for quantitative
purposes. For example, the small stick was less stiff
than the tube, leading to a larger curvature
at the junction. Moreover, the tube undergoes 
plastic deformation,
and it did not come back to a straight shape
after the experiments. Furthermore, solid friction
occurs at the contact of the tube with itself.
Therefore, the curvature energy of the tube may not
be fully relaxed.

\begin{figure}
\centerline{
\hbox{\psfig{figure=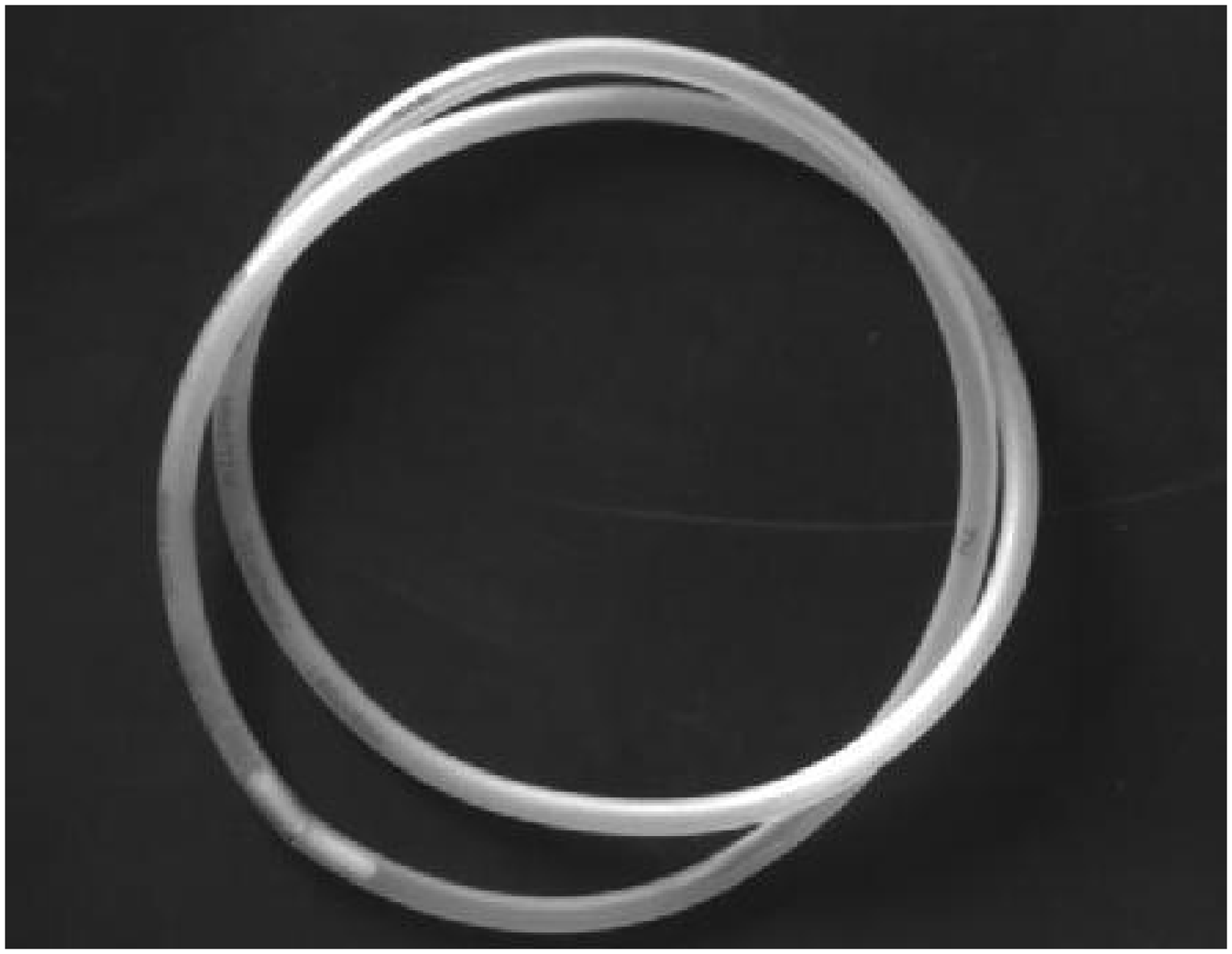,width=4cm,angle=0}}  }
\centerline{
\hbox{\psfig{figure=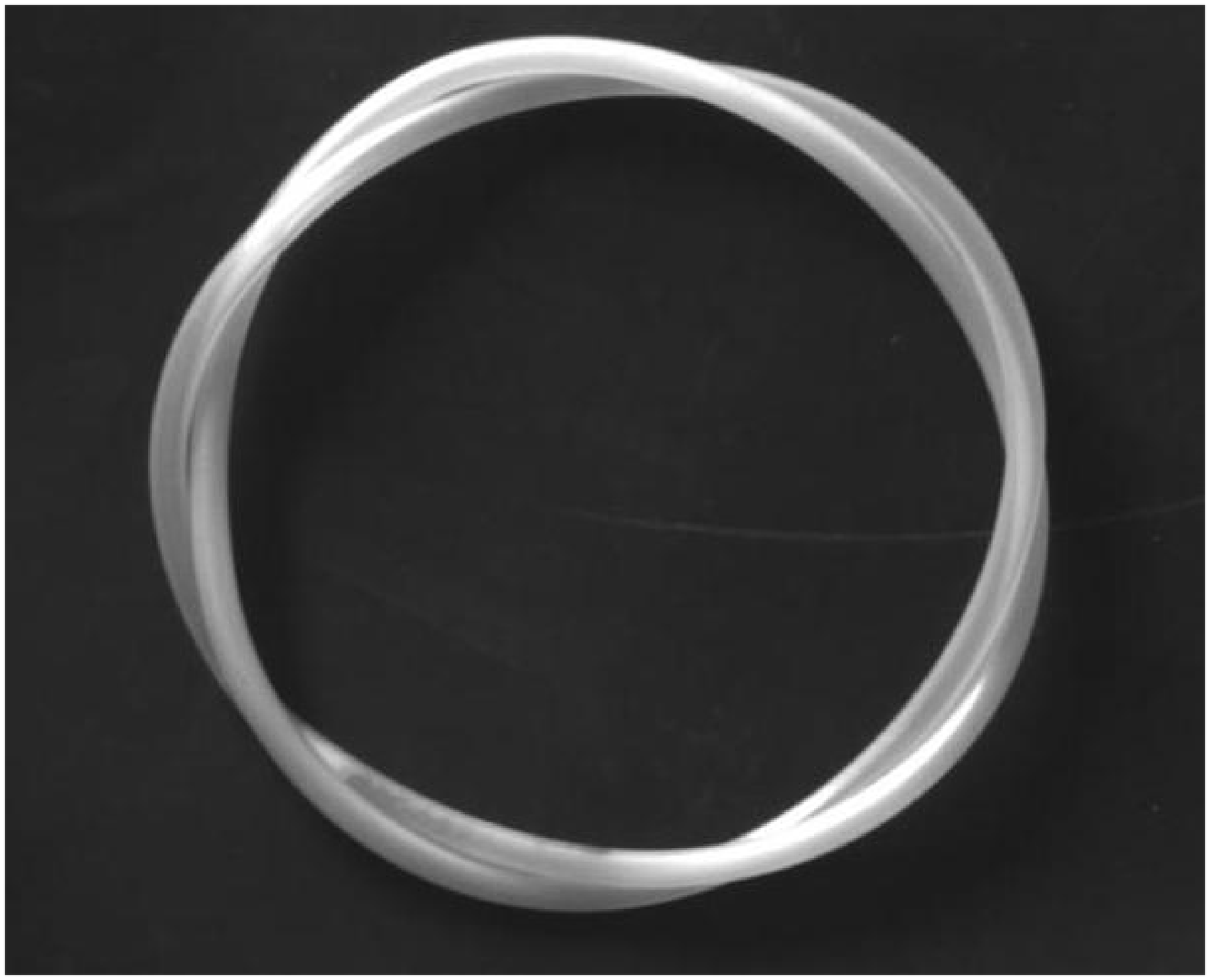,width=4cm,angle=0}}  }
\centerline{
\hbox{\psfig{figure=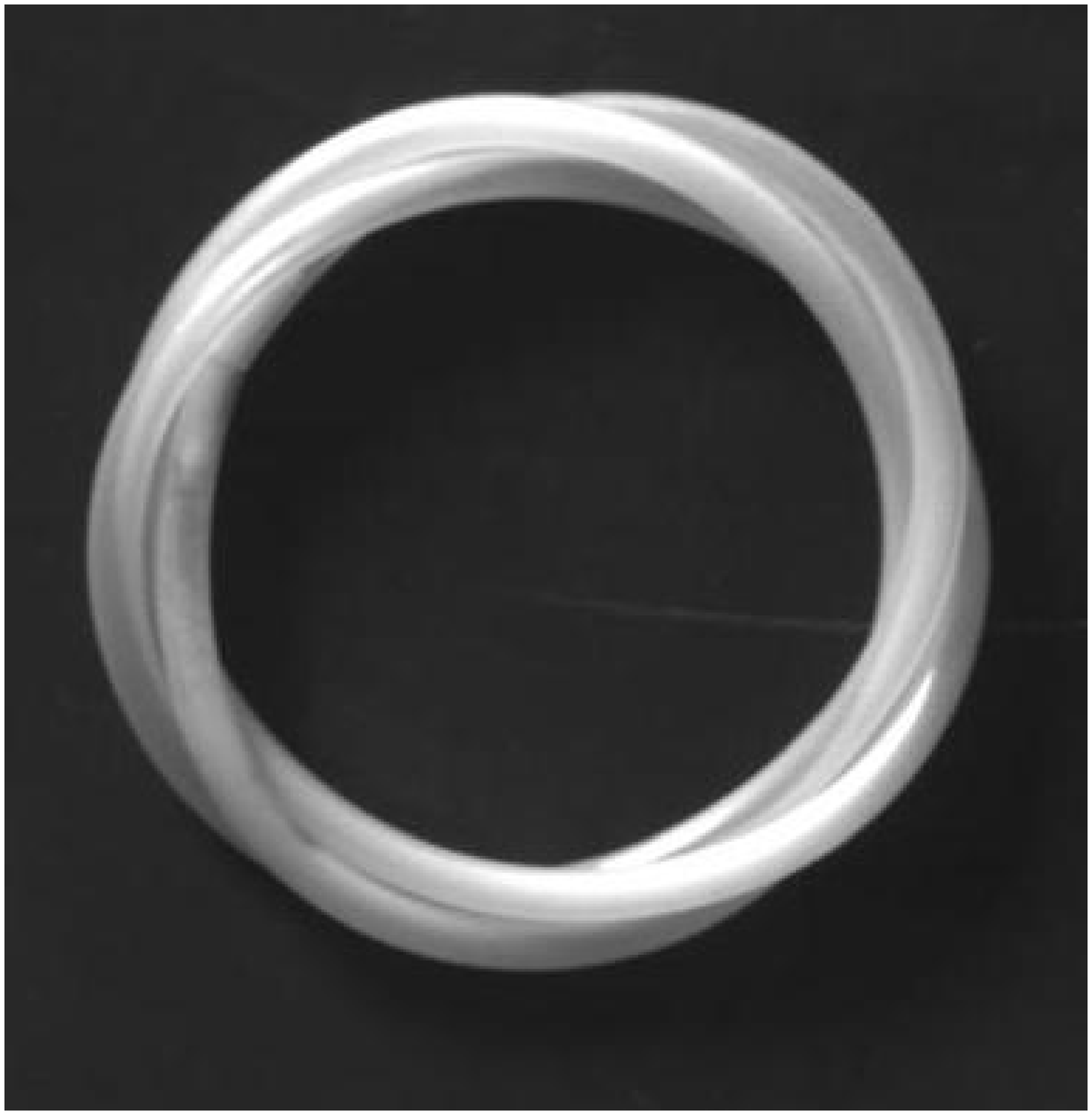,width=4cm,angle=0}}  }
\centerline{
\hbox{\psfig{figure=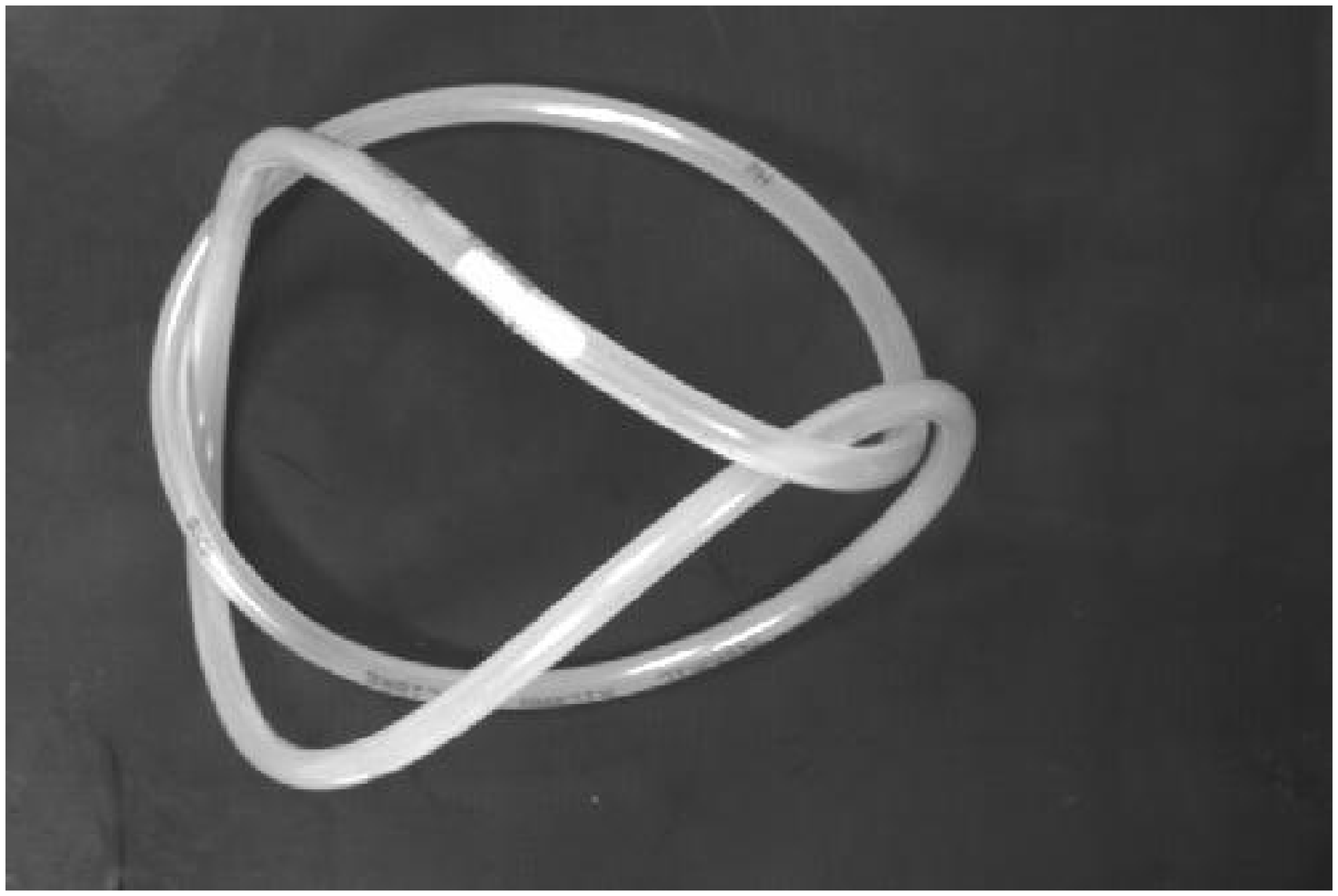,width=4cm,angle=0}}  }
\centerline{
\hbox{\psfig{figure=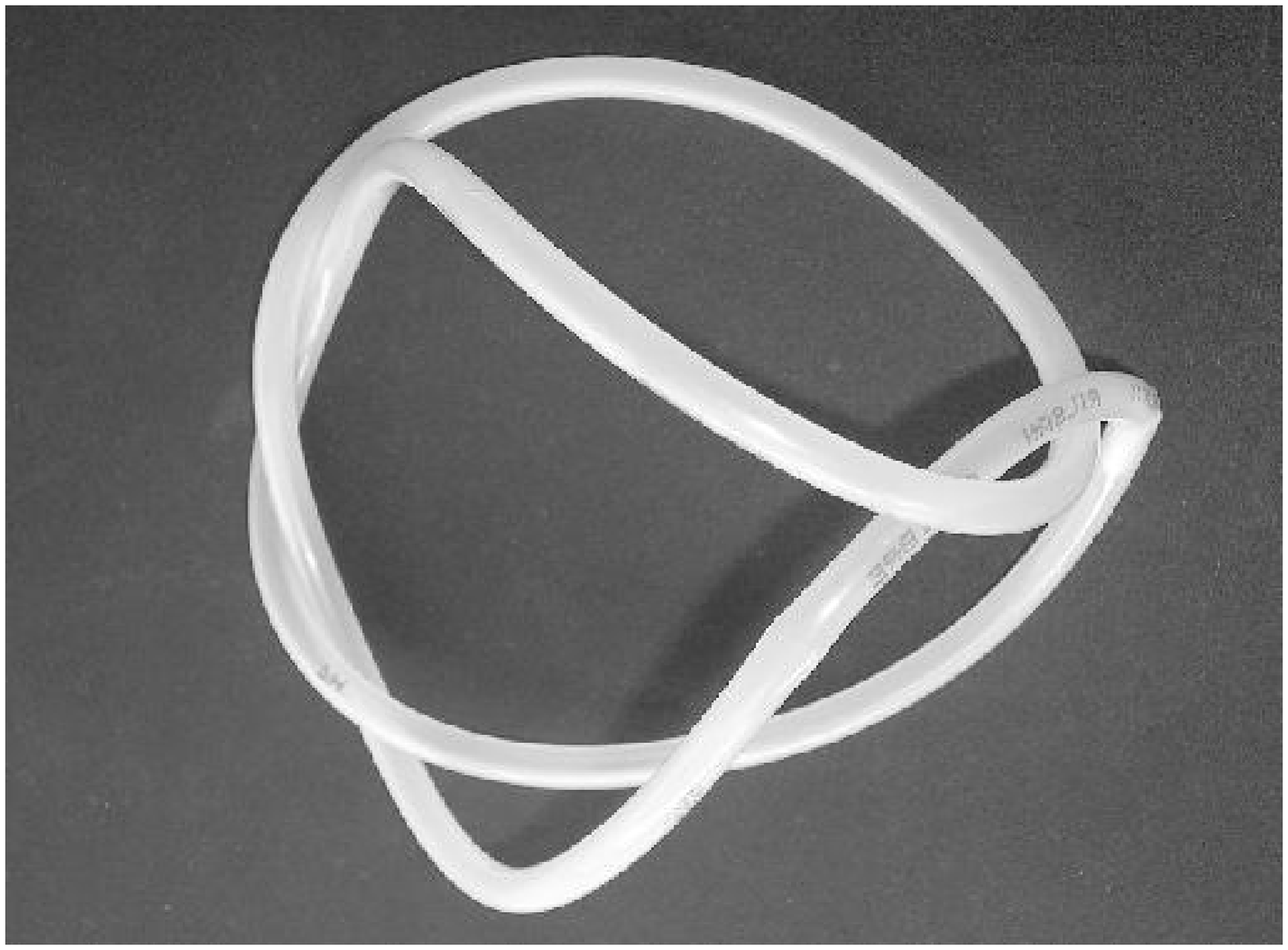,width=4cm,angle=0}}  }
\centerline{
\hbox{\psfig{figure=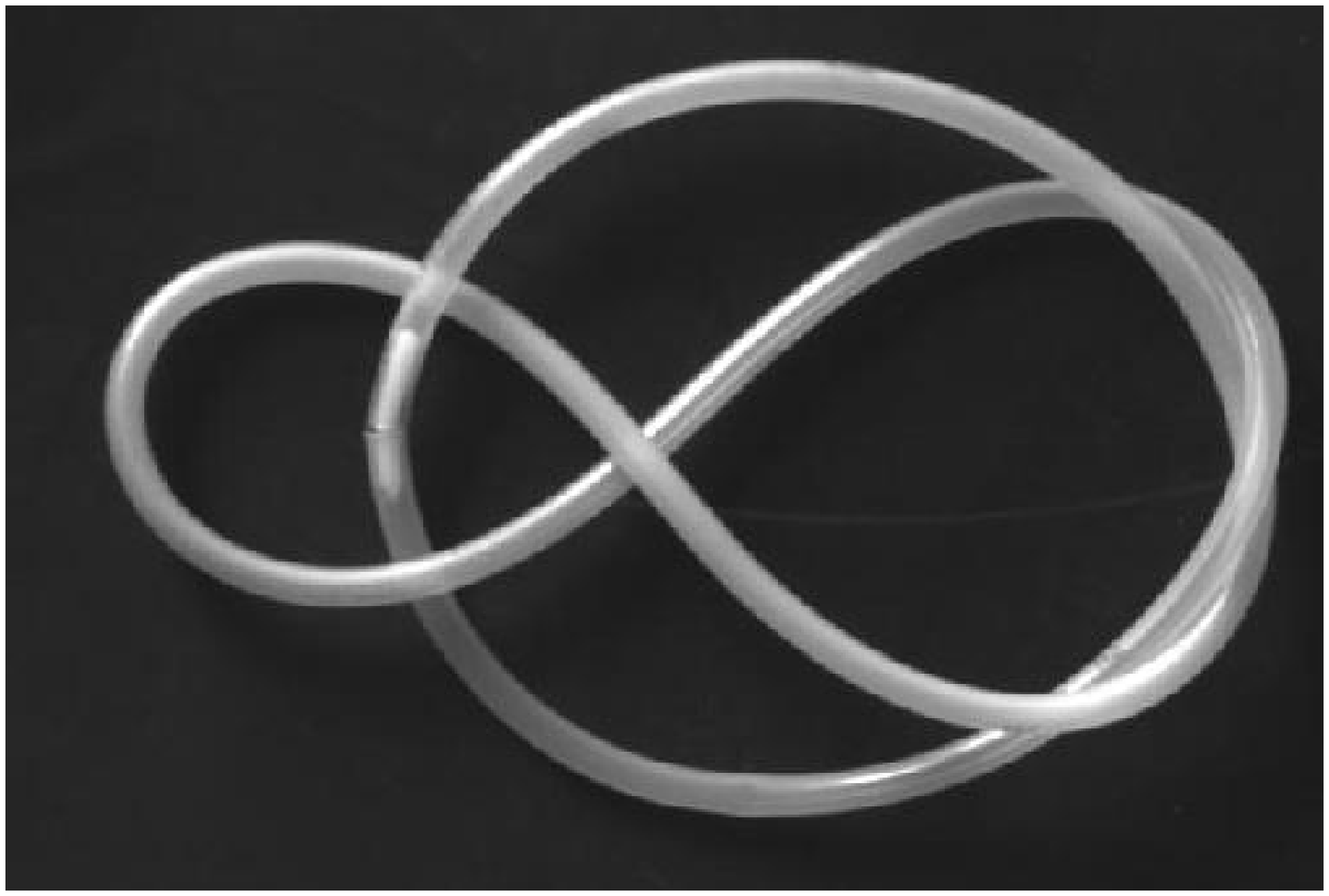,width=4cm,angle=0}}  }
\caption{
Photographs of closed knotted tube:
$3_1$, $5_1$, $T(5,3)$ (torus knot), $4_1$, $5_2$
in 4-fold cage configuration, $5_2$ in 80 configuration.
}
\label{fig:tube_experiments}
\end{figure}

Despite these imperfections, the experiments 
qualitatively confirmed the three
types of shape already obtained, 
as shown on Fig.\ref{fig:tube_experiments}.
Circular braids are obtained
for knots with $n=i$: $3_1$ and $5_1$,
for which $i=2$, and also the $T(5,3)$ torus knot
with $i=3$. 
Moreover, the expected 4-fold cage configuration is obtained
for the $4_1$ and the $5_2$. Finally, 80 configurations is found for
the $5_2$. 
These results are in perfect agreement with the
results of the theory and the
MC simulations of section \ref{s:MC}.

\subsection{The restricted curvature problem}

Other minimization problems share
similarities with the minimization of the
curvature energy ${\cal E}$.
An example is the minimization of the total length
of a knotted filament which exhibits a minimum possible
radius of curvature $R_+$.
This constraint may  result from an excluded
volume effect which limits the bending angle
between  adjacent units in a polymer
or a macroscopic chain. 

Let us assume that curvature $\kappa$ of
a knotted filament cannot exceed $1/R_+$, i.e.
$\kappa\leq 1/R_+$. Our aim here is to
study the  minimum possible length ${\cal L}^\dagger$
of this filament in the limit where $w\rightarrow 0$.
First, from the inequality
$\bar \kappa=\int ds \kappa \leq {\cal L}/R_+$
combined with (\ref{e:bridge_k}),
we obtain a lower bound for the length of a knot with a restricted
curvature:
\begin{eqnarray}
{\cal L}\geq 2\pi nR_+.
\end{eqnarray}
Once again, we want
to avoid angular points and knot localization, 
which exhibit infinite local
curvature $>1/R_+$, and braid localization is
expected in the limit $w\rightarrow 0$.

\begin{figure}
\centerline{
\hbox{\psfig{figure=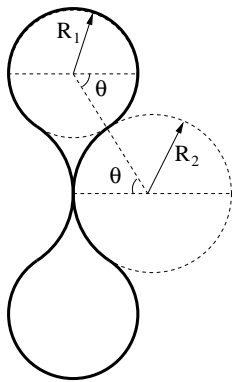,width=3cm,angle=0}}  }
\caption{
Ansatz for the shape of the Point-Braid and Loop (PBL)
configuration. The shape is made of arcs of circles
of radius $R_1$, and $R_2$. The axis passing through
the center of the circles
goes through the contact point between two arcs,
and defines the angle $\theta$.
}
\label{fig:restricted_curvarure_model}
\end{figure}

We then use the same strategy as before
to determine upper bounds.
A first upper bound is found
from the PBL configuration, defined
in Fig.\ref{fig:knot_transformations}.
Because of the discontinuous
character of the constraint $\kappa\leq1/R_+$,
we cannot formulate the shape optimization problem
with differential equations analogous
to Eqs.(\ref{e:var_E_1},\ref{e:var_E_3})
to obtain the minimal shape and energy.
We therefore use a simple ansatz for the geometry,
where loops
are formed by arcs of circles. The precise shape
is defined in Fig.\ref{fig:restricted_curvarure_model}.
The total length is:
\begin{eqnarray}
{\cal L}=2n\left[R_1(\pi+2\theta)+2R_2\theta\right]
\end{eqnarray}
where the free variables are $R_1$, $R_2$.
The constraint of restricted curvature reads:
\begin{eqnarray}
R_1\geq R_+; \;\; R_2\geq R_+\, .
\end{eqnarray}
The angle $\theta$ is given by the relation:
\begin{eqnarray}
R_1\cos(\theta)=R_2(1-\cos(\theta))\, ,
\end{eqnarray}
with $0<\theta<\pi/2$.
The minimization of ${\cal L}$ with the constraints
is straightforward
\footnote{
Defining $\alpha=R_2/(R_1+R_2)$, one has
${\cal L}=2nR_1(\pi+2(1-\alpha)^{-1}{\rm arccos}(\alpha)$.
For fixed $\alpha$, $\partial_{R_1}{\cal L}>0$. Therefore,
the minimum of ${\cal L}$
is reached for $R_1$ equal to its minimum possible value,
i.e. $R_1=R_+$. For fixed $R_1$, $\partial_{\alpha}{\cal L}>0$.
Therefore, $\alpha$  must be equal to its minimum possible value
compatible with $R_1=R_+$, i.e. $\alpha=1/2$. Therefore,
$R_2=R_+$.}
and leads to $R_1=R_2=R_+$, so that $\theta=\pi/3$.
Therefore, the minimum length of the PBL configuration
within our ansatz is:
\begin{eqnarray}
{\cal L}_{PBL}=2nR_+{7 \pi \over 3}
\end{eqnarray}

A second upper bound for ${\cal L}^\dagger$ is deduced
from the circular braid of radius $R_+$, whose length is
$2\pi i R_+$.
Finally, we obtain a result similar to (\ref{e:epsilon_ineq}):
\begin{eqnarray}
2\pi n R_+ \leq {\cal L}^\dagger
\leq 2\pi \min\left[{7\over 3}n;i\right]R_+ \, .
\end{eqnarray}
Similar conclusions are also drawn:
Firstly, We conclude that  the minimum
length scales with the bridge number: ${\cal L}^\dagger\sim nR_+$.
Secondly, the minimization problem is readily solved
when $n=i$: we have ${\cal L}^\dagger=2\pi iR_+$
and the shape is a circular braid.

The restricted curvature problem was recently
addressed numerically in Ref.\cite{Buck2004}
for closed knots.
As expected from the above result,
the circular braid configuration  was found for some torus knots
(with $n=i=2$).
A configuration similar to the 4-fold cage
was also found for the $4_1$ knot.
Whether both problems
always lead to similar geometries is still an open question.

A second result of Ref.\cite{Buck2004} is that
a transition occurs for a finite value of 
the width $w$ of the string.  Does such a transition
exist for stiff knots? Extensive simulations
would be needed in order to answer this question.

\section{Conclusion}
In summary, we find that stiff knot mechanics
crucially depends on the type of knot via
a surprisingly simple quantity:
the bridge number $n$. Indeed both the
equilibrium energy of closed knots and the
equilibrium tension for open knots behave
as $n^2$. Up to now, the open trefoil knot
is the only knot which has been studied
in experiments.
More complex knots have been studied
in the fluctuation dominated regime
for polymers such as DNA \cite{Quake_Vilgis_Katrich}.
We hope that our results will motivate
some numerical and experimental
investigation of the mechanics of more complex 
stiff knots.

As a second central result, we find that braid localization,
which was checked here on simple knots,
is  a general and robust feature of the entanglements
of stiff strings. 
We shall mention two possible consequences of braid localization.
Firstly, the curvature energy
--and thus braid localization-- should be irrelevant
for flexible polymers. But it may be
relevant for entanglements of semi-flexible
polymers and fibers \cite{Morse2001,Rodney2005,
actin-solutions}. Imposing
tangential contacts between the strings, braid localization
questions the usual tube model for polymers, which is
based on a lateral confinement due to simple
crossings\cite{Morse2001}.
Secondly, braid localization also implies localization of friction
and strain variations, and may have some important
consequences on knot-induced polymer and
filament break-up \cite{Saitta1999,actin}.

Our work opens a new line of investigation 
towards the understanding of the geometry
and mechanics of stiff knots. But much yet
remains to be done, and
we would therefore like to conclude with a list of open questions:
(i) In the present work, we have
essentially analyzed
the equilibrium state. But we intuitively
expect the number of metastable states to
increase with the knot complexity.
Can this be quantified?
(ii) The question of the metastable states
naturally leads to a second question: how can we study
local stability (i.e. stability with respect to 
small perturbations)? 
(iii) We have studied the limit $w\rightarrow 0$.
What happens for finite $w$? Here we have
shown that the equilibrium energy ${\cal E}^*_w$
increases with $w$. The results of Buck and Rawdon \cite{Buck2004}
on the similar restricted-curvature problem suggest
that qualitative transitions may occur at finite $w$.
(iv) As shown in basic textbooks such as
Refs.\onlinecite{Doi}, torsion usually plays an important role
in filament mechanics.
How can we include it in the present theory?
We hope to report
along these lines in the near future.

We wish to thank P. Peyla, Y. Colin de Verdi\`ere,
S. Baseilhac, C. Lescop,  M. Eisermann, and H. Meyer
for helpful discussions.
YCdV pointed out the proof of appendix B1.

\begin{appendix}

\section{Monotonic increase of the equilibrium energy with $w$}
In this section, we show that the curvature energy
of a knotted string with tubular excluded volume strictly
decreases as the diameter $w$ of the section of the tube
decreases.

\begin{figure}
\centerline{
\hbox{\psfig{figure=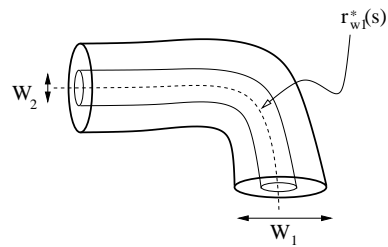,width=5cm,angle=0}}  }
\caption{
Excluded volume tubes of width $w_1$ and $w_2$.
}
\end{figure}

To do so, let us consider a filament denoted as (1), of length
${\cal L}$, and with an excluded volume tube of width $w_1$.
At equilibrium, the central line of the tube (1)
is in the configuration ${\bf r}_1^*(s)$. Let us denote
the energy of this configuration as ${\cal E}^*_{w_1}$.
We then consider another filament of length ${\cal L}$, and
of width $w_2$, with $w_2<w_1$.
If the filament (2) is in the
configuration ${\bf r}_1^*(s)$,
the tube (2) is inside the tube
(1). Therefore, there is no self-intersections
or self-contact for the
tube (2).
Thus,  there is no interactions of the tube
(2) with itself.
Hence, following the results of
section \ref{s:interactions}, ${\bf r}_1^*(s)$
is not the equilibrium configuration for the tube (2).
Therefore, its energy ${\cal E}^*_{w_1}$
in the configuration ${\bf r}_1^*(s)$ is larger than the equilibrium
energy ${\cal E}^*_{w_2}$ of the filament (2).

We have shown that ${\cal E}^*_{w_2}<{\cal E}^*_{w_1}$
when $w_2<w_1$, for any $w_1$ and $w_2$.
As announced in the beginning of this section,
${\cal E}^*$ decreases as $w$ decreases
--or equivalently $\partial_w{\cal E}^*>0$.

\section{Angle and knot localization}
\label{a:angle_knot_localization}

In this appendix, we show that the curvature energy of a curve
in 3D  diverges in the presence of angles,
or knot localization.
To do so,  we will show the equivalent statement that,
in the presence of an upper bound ${\cal E}_0$
for the energy:
\begin{eqnarray}
{\cal E}= {C \over 2}\int ds \kappa^2<{\cal E}_0\, ,
\label{e:upper_bound_e0}
\end{eqnarray}
no angle or knot localization can be present. 

Let us consider a part of the curve ${\bf r}(s)$,
running from  $s_1$ to  $s_2$. From the Schwarz inequality
and (\ref{e:upper_bound_e0}), one has:
\begin{eqnarray}
\int^{s_2}_{s_1} ds\, \kappa
 &\leq& \left(\int^{s_2}_{s_1} ds\right)^{1/2}
 \left(\int^{s_2}_{s_1} ds\, \kappa^2\right)^{1/2}
\nonumber \\
&\leq& 
(s_2-s_1)^{1/2} \left( {2 {\cal E}_0 \over C}\right)^{1/2}
\label{e:local_schwarz}
\end{eqnarray}

\subsection{Angles}
We observe that:
\begin{eqnarray}
|{\bf t}(s_2)-{\bf t}(s_1)|
&=& \left|\int^{s_2}_{s_1} ds\, \partial_s{\bf t}\right|
\nonumber \\
&=& \left|\int^{s_2}_{s_1} ds\, \kappa {\bf n}\right| \leq
\int^{s_2}_{s_1} ds\, \kappa
\label{e:local_tangent_jump}
\end{eqnarray}
where $({\bf t},{\bf n},{\bf b})$ is the usual Frenet
frame along the curve.
Combing
(\ref{e:local_schwarz}) and (\ref{e:local_tangent_jump}),
we obtain:
\begin{eqnarray}
|{\bf t}(s_2)-{\bf t}(s_1)|\leq (s_2-s_1)^{1/2}
\left( {2 {\cal E}_0 \over C}\right)^{1/2}
\label{e:tang_continuity}
\end{eqnarray}
Eq.(\ref{e:tang_continuity}) shows that
the tangent vector ${\bf t}$ is continuous
along the curve. Therefore, no angle can be present.

\subsection{Knot localization}
If the part of the knot
between $s_1$ and $s_2$ is knotted,
then there is a lower
bound for the total curvature which is
similar to Eq.(\ref{e:bridge_k}).
But since the knot is an open one,
one should use the modified
bridge number defined in section \ref{s:open_knots},
which we denote as $n_{1\rightarrow 2}$. We thus have:
\begin{eqnarray}
\int^{s_2}_{s_1} ds\, \kappa \geq
2\pi n_{1\rightarrow 2}
\end{eqnarray}
Using this inequality together with
(\ref{e:local_schwarz}), we obtain:
\begin{eqnarray}
n_{1\rightarrow 2}\leq {1\over 2\pi}
\left({2 {\cal E}_0 \over C}\right)^{1/2} (s_2-s_1)^{1/2}
\label{e:no_knot_loc}
\end{eqnarray}
Knot localization means that there is a
knot between $s_1$ and $s_2$
while $(s_1-s_2)\rightarrow 0$.
But (\ref{e:no_knot_loc}) shows that,
as $(s_2-s_1)\rightarrow 0$,
one necessarily has $n_{1\rightarrow 2}=0$. 
Since $n_{1\rightarrow 2}=0$ corresponds to an unknotted string,
there is no knot localization.

\section{Some relations for a structure of line-braids}
\label{a:variational}

\subsection{Closed knots}
\subsubsection{Relation between ${\cal E}$ and $\mu$}
We here derive a relation between ${\cal E}$, ${\cal L}$, and $\mu$.
Integrating (\ref{e:var_E_1_P}) over the $p$th line-braid,
we obtain:
\begin{eqnarray}
{\cal E}^\circ_p-\mu^\circ m_p{\cal L}^\circ_p={\bf A}_p.[{\bf r}]_p
\label{e:mu_P}
\end{eqnarray}
where $[{\bf r}]_p$ is the difference between ${\bf r}$
at the beginning and at the end of the line-braid.
Summing (\ref{e:mu_P}) over $p$, we obtain:
\begin{eqnarray}
{\cal E}^\circ-\mu^\circ {\cal L}=\sum_{p=1}^N{\bf A}_p.[{\bf r}]_p
=\sum_B {\bf r}_B.\sum_{p\rightarrow B}{\bf A}_p=0
\label{e:sum_mu_closed}
\end{eqnarray}
where the last equality follows from (\ref{e:BC_P_A}).
Finally, one has:
\begin{eqnarray}
\mu ={{\cal E}\over {\cal L} }
\label{e:mu_E_L}
\end{eqnarray}
This equality is true for all structures which obey
(\ref{e:var_E_1_P},\ref{e:var_E_3_P}), with
the boundary conditions (\ref{e:BC_P_A},\ref{e:BC_P_Q}).

\subsubsection{Expression of $\epsilon$}
We now derive a relation between the
total normalized energy, and the normalized
energy of its parts.
Eliminating $\mu$ between Eqs.(\ref{e:mu_P}) and (\ref{e:mu_E_L}),
and rewriting energies in terms of the normalized energies $\epsilon$,
we find
\begin{eqnarray}
m_p\left(\epsilon_p^\circ
-{\bf A}_p.[{\bf r}]_p{{\cal L}_p \over m_p C}\right)^{1/2}
=\epsilon^{\circ 1/2}_{\{{\cal S}\}} {m_p{\cal L}_p \over {\cal L}}
\label{e:normalized_parts}
\end{eqnarray}
where $\epsilon_p={\cal E}_p{\cal L}_p/m_pC$.
Summing over $p$, we get:
\begin{eqnarray}
\epsilon^{\circ 1/2}_{\{{\cal S}\}}=\sum_{p=1}^Nm_p
\left(\epsilon^\circ_p
-{\bf A}_p.[{\bf r}]_p{{\cal L}_p \over m_p C}\right)^{1/2}\, .
\end{eqnarray}
There are three interesting cases in which the second term
in the parenthesis vanishes. The first one is the situation
where line-braids are arcs
of circles, for which ${\bf A}_p=0$. In the second case,
all line-braids are loops (i.e. they start and end at the same
point), implying $[{\bf r}]_p=0$. The third case
is the case where all line-braids have identical energies
and length.
If there is $N$ line-braids,
$\mu={\cal E}/{\cal L}=N{\cal E}_p/Nm_p{\cal L}_p={\cal E}_p/m_p{\cal L}_p$.
Such an equality, combined with
(\ref{e:mu_P}), implies that ${\bf A}_p.[{\bf r}]_p=0$.
In these 3 cases, one finally has:
\begin{eqnarray}
\epsilon_{\{{\cal S}\}}^\circ
= \Bigl[\sum_{p}m_p \epsilon_p^{\circ 1/2}\Bigr]^2 \, ,
\label{e:epsilon*_parts_app}
\end{eqnarray}
and from Eq.(\ref{e:normalized_parts}), one finds
Eq.(\ref{e:L_P_parts}).

\subsection{Open knots}

\subsubsection{Relation between ${\cal E}$ and ${\cal L}_e$}
In the case of open knots,
the relation (\ref{e:mu_P}) is still valid for each
line-braid. But the sum in (\ref{e:sum_mu_closed})
does not vanish, and we obtain:
\begin{eqnarray}
{\cal E}^\circ-\mu^\circ {\cal L}^\circ=\sum_{p=1}^N{\bf A}_p.[{\bf r}]_p
=\sum_B {\bf r}_B.\sum_{p\rightarrow B}{\bf A}_p=[{\bf r}.{\bf A}]_-^+
\label{e:sum_mu_open}
\end{eqnarray}
where the index $\pm$ indicate the two open ends
at $z_\pm\rightarrow \pm\infty$.
Using Eq.(\ref{e:A_P}), we obtain
\begin{eqnarray}
[{\bf r}.{\bf A}]_-^+
\rightarrow -\mu^\circ z_+ +\mu^\circ z_-=-\mu\int dz,
\end{eqnarray}
so that (\ref{e:sum_mu_open}) can be re-written as:
\begin{eqnarray}
\mu= {{\cal E} \over {\cal L}_e} \, ,
\end{eqnarray}
where ${\cal L}_e$ is defined in Eq.(\ref{e:excess_length}).

\subsubsection{Expression of $\epsilon_e$}

Let us denote with the index $+$ and $-$ the
parts of the structure related to the open ends. 
Their excess lengths ($\int ds-\int dz$ along these parts)
are respectively denoted as ${\cal L}_{e+}$ and ${\cal L}_{e-}$.
The $+$ part extends from $z_+\rightarrow +\infty$
to the first contact point, whose abscissa along $z$
is denoted as $z_{0+}$. For the $-$ part,
$z_{0-}$ is defined in a similar way.
We then define $\Delta_0=z_{0+}-z_{0-}$.

Following the same lines as in the previous paragraphs,
one finds that:
\begin{eqnarray}
\epsilon_{e+}^\circ=\epsilon_e^\circ\left({\cal L}_{e+}^\circ\over {\cal L}_e\right)^2
\;\;\;
\epsilon_{e-}^\circ=\epsilon_e^\circ\left({\cal L}_{e-}^\circ\over {\cal L}_e\right)^2
\\
\epsilon_p^\circ=\epsilon_e^\circ\left({\cal L}_{p}^\circ\over {\cal L}_e\right)^2
+ {\bf A}_p.[{\bf r}]_p{{\cal L}_p^\circ \over m_pC}
\end{eqnarray}
Summing these relations over all line-braids lead to:
\begin{eqnarray}
&&\left(1+{\Delta_0 \over {\cal L}_e}\right) \epsilon^{\circ 1/2}_e
\nonumber \\
&=& 
\sum_{p=1}^Nm_p
\left(\epsilon_p^\circ-{\bf A}_p.[{\bf r}]_p{{\cal L}_p
\over m_p C}\right)^{1/2}
+\epsilon_{e+}^{\circ 1/2}+\epsilon_{e-}^{\circ 1/2} \, .
\end{eqnarray}

\section{Energies of special configurations}
\label{a:energy_calc}
\subsection{Energy of the planar loop}
In this appendix, we determine the energy of 
one planar loop in 
the configuration of Fig.\ref{fig:knot_transformations}e.
For planar solutions, ${\bf A}$ is parallel to the plane,
and Eqs.(\ref{e:var_E_1},\ref{e:var_E_3}) reduce to:
\begin{eqnarray}
{1 \over 2}(\partial_s\theta)^2
-{1 \over C}(\mu+A\cos\theta)=0 \, , 
\label{e:2D_elastica}
\end{eqnarray}
where $A=|{\bf A}|$, and $\theta$ is the angle
between the tangent vector ${\bf t}$ and ${\bf A}$.

\begin{figure}
\centerline{
\hbox{\psfig{figure=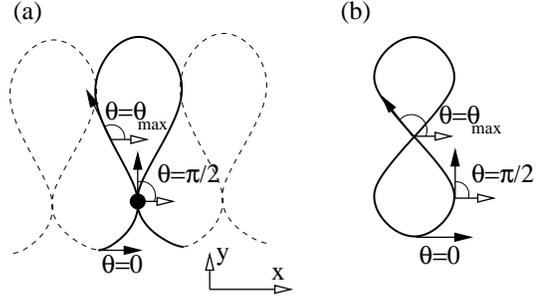,width=7cm,angle=0}}  }
\caption{
Schematics of (a) the loop solution. 
The loop part of solution used in the text starts and ends
at the place indicated by the black dot.
(b) the 8 solution.
}
\label{fig:energy_calc}
\end{figure}

The abscissa $x$, defined in Fig.\ref{fig:energy_calc},  
is written as:
\begin{eqnarray}
x=\int^x_0 dx =\int_0^\theta d\theta {\partial_sx \over \partial_s\theta}
=\int_0^\theta d\theta {\cos\theta \over \partial_s\theta} \, ,
\end{eqnarray}
where $\partial_s\theta$ is given by Eq.(\ref{e:2D_elastica}).
Using the variable change
$v=-\alpha\cos\theta$, with $\alpha=A/\mu$, we find:
\begin{eqnarray}
x=\alpha^{-1} \left(C \over \mu\right)^{1/2}
\int_v^\alpha {v\, dv \over
(\alpha^2-v^2)^{1/2}(v+1)^{1/2}} \, .
\end{eqnarray}
As seen from an inspection of Fig.\ref{fig:energy_calc}a,
the constraint which selects the loop solution with
a tangent vector along $y$ at the boundary is:
\begin{eqnarray}
x(\theta=\pi/2)-x(\theta=0)=2(x(\theta=\pi/2)-x(\theta_{max})),
\end{eqnarray}
where $\theta_{max}$ is such that $v_{max}=-1$.
This constraint leads to an equation for $\alpha$:
\begin{eqnarray}
2\int_{-1}^0 {v\, dv \over
(\alpha^2-v^2)^{1/2}(v+1)^{1/2}}
\nonumber \\
+\int_0^\alpha {v\, dv \over
(\alpha^2-v^2)^{1/2}(v+1)^{1/2}}=0 \, .
\label{e:alpha_loop}
\end{eqnarray}

Moreover, since $\kappa=|\partial_s\theta|$, the energy reads:
\begin{eqnarray}
{\cal E}_{loop}={C \over 2}\int ds (\partial_s\theta)^2\, ,
\end{eqnarray}
and the total length is:
\begin{eqnarray}
{\cal L}_{loop}=\int ds\, .
\end{eqnarray}
These integrals are both re-written with the 
new variable $v$, and finally the normalized energy reads:
\begin{eqnarray}
\epsilon_{loop}^\circ&=&{{\cal E}_{loop}{\cal L}_{loop}\over C}
\nonumber \\
&=&2\left[2\int_{-1}^0  dv 
{(v+1)^{1/2} \over (\alpha^2-v^2)^{1/2}}
+\int_0^\alpha  dw 
{(v+1)^{1/2} \over (\alpha^2-v^2)^{1/2}}\right]
\nonumber \\
&&\times
\left[2\int_{-1}^0 { dv \over
(\alpha^2-v^2)^{1/2}(v+1)^{1/2}}\right.
\nonumber \\
&& \hspace{0.3 cm}\left.+\int_0^\alpha { dv \over
(\alpha^2-v^2)^{1/2}(v+1)^{1/2}}\right]\, .
\label{e:epsilon_loop_calc}
\end{eqnarray}
The numerical solution of Eq.(\ref{e:alpha_loop})
leads to $\alpha\approx2.158$. Substituting
this value in Eq.(\ref{e:epsilon_loop_calc}), we find
$\epsilon_{loop}^\circ\approx18.19$.

\subsection{Energy of the 8}
For the '8', we use the same method as in
the previous section.
The constraint is now the periodicity of the curve,
which imposes 
\begin{eqnarray}
x(\theta_{max})=x(-\theta_{max})\, ,
\end{eqnarray}
where $\theta_{max}$ is the maximum angle
reached along the curve. 
This leads to the following condition
after the change of variable to $v$:
\begin{eqnarray}
\int_{-1}^\alpha {v\, dv \over
(\alpha^2-v^2)^{1/2}(v+1)^{1/2}}=0 \, ,
\end{eqnarray}
which is solved numerically, and leads to $\alpha=1.53$.
Moreover,
\begin{eqnarray}
\epsilon_8^\circ
&=&{{\cal E}_8{\cal L}_8\over C}
\nonumber \\
&=&8\left[\int_{-1}^\alpha  dv
{(v+1)^{1/2} \over (\alpha^2-v^2)^{1/2}}\right]
\nonumber \\
&& \times
\left[\int_{-1}^\alpha { dv \over
(\alpha^2-v^2)^{1/2}(v+1)^{1/2}}\right]\, .
\end{eqnarray}
We find numerically that $\epsilon_8\approx 55.93$.

\end{appendix}

\end{document}